# Efficient Uncertainty Quantification for Dynamic Subsurface Flow with Surrogate by Theory-guided Neural Network

Nanzhe Wang[a], Haibin Chang[a], and Dongxiao Zhang[b,c,*]

[a] BIC-ESAT, ERE, and SKLTCS, College of Engineering, Peking University, Beijing 100871, P. R. China
[b] School of Environmental Science and Engineering, Southern University of Science and Technology, Shenzhen 518055, P. R. China
[c] Intelligent Energy Lab, Peng Cheng Laboratory, Shenzhen 518000, P. R. China

[*] Corresponding author: E-mail address: zhangdx@sustech.edu.cn (Dongxiao Zhang)

**Abstract**

Subsurface flow problems usually involve some degree of uncertainty. Consequently, uncertainty quantification is commonly necessary for subsurface flow prediction. In this work, we propose a methodology for efficient uncertainty quantification for dynamic subsurface flow with a surrogate constructed by the Theory-guided Neural Network (TgNN). The TgNN here is specially designed for problems with stochastic parameters. In the TgNN, stochastic parameters, time and location comprise the input of the neural network, while the quantity of interest is the output. The neural network is trained with available simulation data, while being simultaneously guided by theory (e.g., the governing equation, boundary conditions, initial conditions, etc.) of the underlying problem. The trained neural network can predict solutions of subsurface flow problems with new stochastic parameters. With the TgNN surrogate, the Monte Carlo (MC) method can be efficiently implemented for uncertainty quantification. The proposed methodology is evaluated with two-dimensional dynamic saturated flow problems in porous medium. Numerical results show that the TgNN based surrogate can significantly improve the efficiency of uncertainty quantification tasks compared with simulation based implementation. Further investigations regarding stochastic fields with smaller correlation length, larger variance, changing boundary values and out-of-distribution variances are performed, and satisfactory results are obtained.

**Keywords:** Theory-guided Neural Network; surrogate modeling; subsurface flow; uncertainty quantification.

## 1. Introduction

Subsurface flow problems usually involve some degree of uncertainty, resulting from heterogeneity of porous medium and incomplete information about the medium's geological properties. Therefore, quantifying uncertainty of the subsurface flow response induced from the media's properties is usually necessary (Smith, 2013; Zhang,

2001). The Monte Carlo (MC) method is the most straightforward way to perform uncertainty quantification (UQ), which can estimate the statistical properties of model responses via random sampling. While using the MC method, a large number of realizations of stochastic input should be generated and evaluated to guarantee accuracy of the estimated statistical quantities (Ballio & Guadagnini, 2004). As a result, the computational cost may be very large, especially for large-scale problems, which constitutes the major drawback of the MC method.

In order to overcome the high computational cost of the MC method, the surrogate modeling technique is frequently utilized. For constructing a surrogate, a series of realizations should be generated and evaluated to provide training data. The constructed surrogate model can approximate the relationship between the inputs and outputs of subsurface flow problems. In addition, forward evaluation by using the constructed surrogate requires much lower computational cost than that by running the simulator directly. By using the surrogate model instead of the simulator, under certain conditions, efficiency will be significantly improved when performing UQ tasks with the MC method. Some widely utilized surrogate models include Gaussian Process (GP) (Kennedy & O'Hagan, 2000; Williams & Rasmussen, 2006), Polynomial Chaos Expansion (PCE) (Chang & Zhang, 2009; Li & Zhang, 2007; Liao & Zhang, 2015; Xiu & Karniadakis, 2002a, 2002b), and radial basis functions (Park & Sandberg, 1991; Regis & Shoemaker, 2007).

Although the aforementioned traditional surrogate models have achieved great progress, it remains challenging to tackle the high-dimensional problems for those methods, owing to the 'curse of dimensionality'. Indeed, for problems with high dimensionality, the computational cost for constructing a surrogate model may be extremely large. Besides these surrogate models, the Deep Neural Network (DNN) has received increasing attention because of advancements in computer hardware technology, such as Graphics Processing Units (GPUs), which provides powerful computing resources and speeds up calculations. DNN has also been used for surrogate modeling due to its universal approximation ability and great potential to deal with high-dimensional nonlinear problems. Tripathy and Bilionis (2018) proposed a DNN-based surrogate, which was used for UQ tasks of a high-dimensional stochastic elliptic partial differential equation (SPDE) with uncertain diffusion coefficient. Zhu and Zabaras (2018) developed a surrogate based on a deep convolutional encoder-decoder network, which transformed the output prediction into an image-to-image regression task. The proposed surrogate was also tested with UQ tasks of SPDEs with high-dimensional stochastic input. Mo et al. (2019) applied a deep autoregressive neural network-based surrogate model, which employed the convolutional encoder-decoder network architecture, to solve inversion problems, such as groundwater contaminant source identification. Moreover, Mo et al. (2019) developed a convolutional encoder-decoder network based surrogate model, which was trained via combined regression loss and segmentation loss to better characterize the discontinuous saturation front. UQ for dynamics multiphase flow was performed via the proposed surrogate.

Despite the numerous successes achieved with DNN surrogates, some limitations still exist for this purely data-fitting method. First, a large number of training data are usually required to build a DNN surrogate. Usually, hundreds or thousands of simulation runs are needed to provide adequate labeled data for neural network training, which brings about large computation cost. Second, physical laws and scientific knowledge are not considered when building the DNN surrogates, which are only driven by data-fitting. The DNN surrogates may provide unreasonable outputs without obeying physical laws or being guided by theory during the training process. To transcend these limitations, incorporating physical constraints as prior knowledge into the neural network training has been investigated in some works. Raissi et al. (2019) proposed a Physics-Informed Neural Network (PINN), which incorporates the Partial Differential Equations (PDEs) residual into the loss function as a regularization term. The PINN can be used to solve PDEs and identify (a small number of) unknown parameters in PDEs. Wang et al. (2020) developed a Theory-guided Neural Network (TgNN), in which not only physics principles, but also practical engineering theories (e.g., engineering controls and expert knowledge), are incorporated into the neural network training. The physics-constrained neural network has recently received increasing focus and been utilized for surrogate modeling. A physics-constrained convolutional encoder-decoder neural network surrogate was proposed by Zhu et al. (2019). The proposed surrogate was trained without labeled data, and the Sobel filter was adopted to estimate the spatial gradients of the Convolutional Neural Network (CNN) architecture. While approximating the time gradient in two-dimensional (2D) subsurface flow problems with the Sobel filter is difficult, only steady-state flows are considered in their work. Karumuri et al. (2020) developed a residual Fully Connected Neural Network (FCNN)-based surrogate, which is also trained without labeled data. In their work, the uncertain parameters of each grid blocks are inputted into the network, and only steady-state problems are investigated. Sun et al. (2020) applied the physics-constrained neural network for surrogate modeling of fluid flow problems, which are governed by Navier–Stokes equations. In their work, only steady-state cases are considered for proof-of-concept, and uncertain parameters are relatively limited (e.g., viscosity, tube curvature). Although constructing a DNN surrogate with simulator-free or data-free training has gained increasing attention, whether this new strategy is superior to constructing surrogates with the aid of data remains undetermined.

In this work, we proposed a methodology for efficient uncertainty quantification for dynamic subsurface flow with a surrogate constructed by the Theory-guided Neural Network (TgNN). The TgNN here is specially designed for problems with stochastic parameters. In the TgNN, stochastic parameters, time and location comprise the input of the neural network, while the quantity of interest is the output. Similar to Wang et al. (2020), the neural network is trained with available simulation data, while being simultaneously guided by theory (e.g., the governing equation, boundary conditions, initial conditions, etc.) of the underlying problem. The trained neural network can predict solutions of subsurface flow problems with new stochastic parameters. The MC method is then adopted for UQ with assistance from the TgNN surrogate. The proposed methodology is evaluated with 2D dynamic saturated flow problems in porous medium. Numerical results demonstrate that the TgNN based surrogate can significantly improve the efficiency of UQ tasks compared with simulation based implementation. The contribution of our work comprises the following aspects. First, different from previous physics-constrained surrogate works in which only steady state problems are investigated, in our work, dynamical problems are studied. Second, for the deep learning-based surrogate, the effect of labeled-data volume and the number of collocation points that are used for enforcing physical law constraints are thoroughly examined, in order to assess the performance of the labeled-data based and label-free learning. Third, problems with varying boundary conditions and field variances are investigated, and a composited surrogate model is proposed. Fourth, the issue of limited extrapolation capacity of the composited surrogate is examined, and a transfer learning-based strategy is proposed, which can expand applicability of the proposed surrogate.

The remainder of this paper is organized as follows. In section 2, the considered subsurface flow problem, the KLE-based parameterization, and the TgNN surrogate based UQ are briefly introduced. In section 3, the performance of the TgNN surrogate based UQ is tested with dynamical subsurface flow cases in various scenarios. Finally, the summary of the work is presented in section 4.

## 2. Methodology

### 2.1 Subsurface flow problem

In this work, we consider a transient saturated subsurface flow problem with a general form of governing equation:

$$S_s \frac{\partial h(\mathbf{x},t)}{\partial t} - \nabla \cdot (K(\mathbf{x})\nabla h(\mathbf{x},t)) = 0 \tag{1}$$

subject to boundary conditions:

$$h(\mathbf{x},t) = h_D(\mathbf{x}), \quad \mathbf{x} \in \Gamma_D \tag{2}$$

$$K(\mathbf{x})\nabla h(\mathbf{x}) \cdot \mathbf{n}(\mathbf{x}) = g(\mathbf{x}), \quad \mathbf{x} \in \Gamma_N \tag{3}$$

$$h(\mathbf{x},0) = h_0(\mathbf{x}) \tag{4}$$

where $S_s$ denotes the specific storage; $h(\mathbf{x},t)$ denotes the hydraulic head; $K(\mathbf{x})$ denotes the hydraulic conductivity; $h_D(\mathbf{x})$ denotes the prescribed head on Dirichlet boundary segments $\Gamma_D$; $g(\mathbf{x})$ denotes the prescribed flux across Neumann boundary segments $\Gamma_N$; $\mathbf{n}(\mathbf{x})$ denotes an outward unit vector normal to the boundary; and $h_0(\mathbf{x})$ denotes the hydraulic head at $t=0$, i.e., the initial condition.

### 2.2 Parameterization with Karhunen–Loeve expansion (KLE)

The parameter fields of subsurface flow problems are commonly heterogeneous,

and large uncertainties about these fields usually exist due to limited information. Therefore, in this work, the hydraulic conductivity field $K(\mathbf{x})$ is treated as a random field. Although other dimension reduction techniques, such as Singular Value Decomposition (SVD) (Tavakoli & Reynolds, 2011), discrete wavelet transform (Awotunde & Horne, 2013), discrete cosine transform (Jafarpour et al., 2010), and auto-encoder (Wang et al., 2016) may be utilized, Karhunen-Loeve expansion (KLE) is adopted in this work for parameterizing the considered random field, which can honor two-point spatial statistics.

For a random field $Z(\mathbf{x},\tau) = \ln K(\mathbf{x},\tau)$, where $\mathbf{x} \in D$ (physical domain) and $\tau \in \Theta$ (probability space), it can be expressed as $Z(\mathbf{x},\tau) = \bar{Z}(\mathbf{x}) + Z'(\mathbf{x},\tau)$, where $\bar{Z}(\mathbf{x})$ is the mean of the random field, and $Z'(\mathbf{x},\theta)$ is the fluctuation. The spatial structure of the random field can be described by the two-point covariance $C_Z(\mathbf{x},\mathbf{x}') = \langle Z'(\mathbf{x},\tau)Z'(\mathbf{x}',\tau) \rangle$. Using KLE, $Z(\mathbf{x},\tau)$ can be expressed as (Ghanem & Spanos, 2003):

$$Z(\mathbf{x},\tau) = \bar{Z}(\mathbf{x}) + \sum_{i=1}^{\infty} \sqrt{\lambda_i} f_i(\mathbf{x}) \xi_i(\tau) \qquad (5)$$

where $\lambda_i$ and $f_i(\mathbf{x})$ are the eigenvalue and eigenfunction of the covariance, respectively; and $\xi_i(\tau)$ are orthogonal Gaussian random variables with zero mean and unit variance if $Z(\mathbf{x},\tau)$ is a Gaussian random field, and the KLE can achieve a mean square convergence (Ghanem & Spanos, 2003). For the non-Gaussian field, Kernel Principle Component Analysis (KPCA) can be applied for parameterization (Li & Zhang, 2013).

In particular, for separable exponential covariance function $C_Z(\mathbf{x},\mathbf{x}') = \sigma_Z^2 \exp(-|x_1-x_2|/\eta_x - |y_1-y_2|/\eta_y)$, where $\sigma_Z^2$ and $\eta$ are the variance and the correlation length of the random field, respectively, the eigenvalues and eigenfunctions can be solved analytically or semi-analytically, additional details of which can be found in Zhang and Lu (2004).

There are infinite terms in Eq.(5), and one may truncate the expansion with a finite number of terms ($n$). The number of retained terms in the KL expansion should be determined by the decay rate of $\lambda_i$. Moreover, the number of retained terms ($n$) determines the random dimensionality. By using the KLE, the random field $Z(\mathbf{x},\tau) = \ln K(\mathbf{x},\tau)$ can be parameterized by a group of independent random variables

as follows:

$$\xi = \{\xi_1(\tau), \xi_2(\tau), \cdots, \xi_n(\tau)\} \tag{6}$$

The random field can then be represented as:

$$Z(x,y) \approx \overline{Z}(x,y) + \sum_{i=1}^{n} \sqrt{\lambda_i} f_i(x,y) \xi_i(\tau) \tag{7}$$

**2.3 Theory-guided Neural Network (TgNN) surrogate**

*2.3.1 Deep Neural Network*

The Deep Neural Network (DNN) is a powerful function approximator, which can learn the relationship between model outputs and model inputs after training with adequate data. Various architectures of DNN have been proposed for performing various tasks, such as the Convolutional Neural Network (CNN), the Recurrent Neural Network (RNN), the Generative Adversarial Network (GAN), etc. In this work, a deep Fully-Connected Neural Network (FCNN) architecture is adopted.

There is an input layer, an output layer, and hidden layers in the FCNN architecture, each of which consists of a number of neurons. A DNN usually has more than one hidden layer. The output of each layer serves as input of the next layer, and the forward formulation of the $i^{th}$ hidden layer is given as:

$$\mathbf{z}^i = \sigma_i(\mathbf{W}^i \mathbf{z}^{i-1} + \mathbf{b}^i) \tag{8}$$

where $\mathbf{W}^i$ and $\mathbf{b}^i$ are weights and bias of the $i^{th}$ layer, respectively, which are known as network parameters, $\theta = \{\mathbf{W}^i, \mathbf{b}^i\}_{i=1}^{L+1}$ (here, assume that there are $L$ hidden layers and superscript $L+1$ denotes the output layer); and $\sigma^i$ is the activation function of the $i^{th}$ layer, such as Sigmoid, hyperbolic tangent (Tanh), and Rectified Linear Unit (ReLU) (Goodfellow et al., 2016). In this work, a 'swish' activation function is employed (Ramachandran et al., 2017), which is defined as follows:

$$\sigma(\mathbf{z}) = \frac{\mathbf{z}}{1 + \exp(-\beta \mathbf{z})} \tag{9}$$

where $\beta$ can be either a pre-defined constant or a trainable parameter during the training process. $\beta$ is set to be 1 in this work.

In this work, the quantity of interest is the hydraulic head $(h)$. Since conductivity is treated as a random field and depends on random vector $\xi$, the hydraulic head is also random and can be expressed as $h(t, \mathbf{x}, \xi)$. In order to efficiently quantify the uncertainty of $h(t, \mathbf{x}, \xi)$, DNN is utilized to build a surrogate. Furthermore, the time, location, and stochastic parameters comprise the input of the neural network, i.e., $(t, \mathbf{x}, \boldsymbol{\xi})$. Consequently, the forward formulation of the network can be simply expressed as $\hat{h}(t, \mathbf{x}, \boldsymbol{\xi}) = NN(t, \mathbf{x}, \boldsymbol{\xi}; \theta)$, where $\hat{h}$ denotes the predicted hydraulic head. Then, the loss function, which is usually the mean square error between the predicted and the ground truth data, can be represented as:

$$L(\theta) = MSE_{DATA} = \frac{1}{N}\sum_{i=1}^{N}\left|\hat{h}_i - h_i\right|^2 = \frac{1}{N}\sum_{i=1}^{N}\left|NN(t_i, \mathbf{x}_i, \boldsymbol{\xi}_i; \theta) - h_i\right|^2 \tag{10}$$

where $N$ denotes the total number of labeled data. The labeled data can be obtained from numerical simulations. The neural network can be trained via optimization algorithms, such as stochastic gradient descent (Bottou, 2010), to tune the network parameters and minimize the loss function. The abovementioned procedure is purely data-driven, and abundant data are usually required. However, generating adequate data from simulation for DNN construction may be time-consuming. In addition, the DNN surrogates may provide unreasonable outputs without obeying physical laws or being guided by theory during the training process.

*2.3.2 Theory-guided training for neural network*

To overcome the limitations of DNN, incorporating physical/engineering constraints as prior knowledge into the neural network training is a reasonable option. Following the ideas of the Physics-Informed Neural Network (PINN) (Raissi et al., 2019) and the Theory-guided Neural Network (TgNN) (Wang et al., 2020), a neural network that is trained with available data, while simultaneously adhering to physical laws and engineering theories, can serve as a surrogate for uncertainty quantification (UQ). In order to achieve theory-guided training, the governing equations, boundary conditions, and initial conditions are incorporated into the loss function to guide the training process of the network, instead of purely relying on data. The DNN approximation $\hat{h}(t, \mathbf{x}, \boldsymbol{\xi}) = NN(t, \mathbf{x}, \boldsymbol{\xi}; \theta)$ can be substituted into the governing equation Eq.(1), and the residual can be expressed as:

$$f = S_s \frac{\partial NN(t, \mathbf{x}, \boldsymbol{\xi}; \theta)}{\partial t} - \nabla_\mathbf{x} \cdot \left[K(\mathbf{x})\nabla_\mathbf{x} NN(t, \mathbf{x}, \boldsymbol{\xi}; \theta)\right] \tag{11}$$

Substituting the KLE of $K(\mathbf{x})$ into Eq.(11) yields:

$$f = S_s \frac{\partial NN(t, \mathbf{x}, \boldsymbol{\xi}; \theta)}{\partial t} - \nabla_\mathbf{x} \cdot \left\{\exp\left[\overline{Z}(\mathbf{x}) + \sum_{i=1}^{n}\sqrt{\lambda_i}f_i(\mathbf{x})\xi_i(\tau)\right]\nabla_\mathbf{x} NN(t, \mathbf{x}, \boldsymbol{\xi}; \theta)\right\} \tag{12}$$

The residual then constitutes the physics-constrained loss as follows:

$$MSE_{PDE} = \frac{1}{N_c}\sum_{i=1}^{N_c}\left|f(t_i, \mathbf{x}_i, \boldsymbol{\xi}_i; \theta)\right|^2 \tag{13}$$

where $\{t_i, \mathbf{x}_i, \boldsymbol{\xi}_i\}_{i=1}^{N_c}$ denotes the collocation points; and $N_c$ denotes the total number of collocation points (Raissi et al., 2019). The collocation points are the places where the physical constraints are imposed in high-dimensional parameter space, and can be randomly chosen in the space because no labels are needed for these points. Moreover, the boundary and initial conditions of the PDE system can also be imposed as follows:

$$MSE_D = \frac{1}{N_D} \sum_{i=1}^{N_D} \left| NN(t_i^d, \mathbf{x}_i^d, \xi_i^d; \theta) - h_D(\mathbf{x}_i^d) \right|^2 \quad (14)$$

$$MSE_N = \frac{1}{N_N} \sum_{i=1}^{N_N} \left| K(\mathbf{x}_i^n) \nabla_\mathbf{x} NN(t_i^n, \mathbf{x}_i^n, \xi_i^n; \theta) \cdot \mathbf{n}(\mathbf{x}_i^n) - g(\mathbf{x}_i^n) \right|^2 \quad (15)$$

$$MSE_I = \frac{1}{N_I} \sum_{i=1}^{N_I} \left| NN(0, \mathbf{x}_i^0, \xi_i^0; \theta) - h_0(\mathbf{x}_i^0) \right|^2 \quad (16)$$

where $\{t_i^d, \mathbf{x}_i^d, \xi_i^d\}_{i=1}^{N_D}$ denotes the collocation points at the Dirichlet boundary segments $\Gamma_D$; $\{t_i^n, \mathbf{x}_i^n, \xi_i^n\}_{i=1}^{N_N}$ denotes the collocation points at the Neumann boundary segments $\Gamma_N$; and $\{0, \mathbf{x}_i^0, \xi_i^0\}_{i=1}^{N_I}$ denotes the collocation points at the initial time. This constitutes the 'soft way' to impose the boundary and initial conditions as described above.

Therefore, the total loss function of the theory-guided training can be written as:

$$L(\theta) = \lambda_{DATA} MSE_{DATA} + \lambda_{PDE} MSE_{PDE} + \lambda_D MSE_D \\ + \lambda_N MSE_N + \lambda_I MSE_I \quad (17)$$

where $\lambda_{DATA}$, $\lambda_{PDE}$, $\lambda_D$, $\lambda_N$ and $\lambda_I$ are the hyper-parameters, which control the weight of each term in the loss function. In addition to the abovementioned constraint terms, when other theories are available, such as expert knowledge and engineering control, they can be incorporated in a similar manner (Wang et al., 2020). Then, the neural network can be trained by minimizing the loss function via some optimization algorithms, such as Stochastic Gradient Descent (SGD), Adagrad, Adaptive Moment Estimation (Adam) (Kingma & Ba, 2015), etc. Once trained, this TgNN can work as a surrogate model, the framework of which is shown in **Fig. 1.** For any input $(t, \mathbf{x}, \xi)$, the TgNN surrogate can easily provide the prediction $\hat{h}(t, \mathbf{x}, \xi)$ with no simulation required. Here, it is worth mentioning that the input is not restricted to $(t, \mathbf{x}, \xi)$, i.e., other factors can also be taken into consideration, such as uncertain field variance and boundary values.

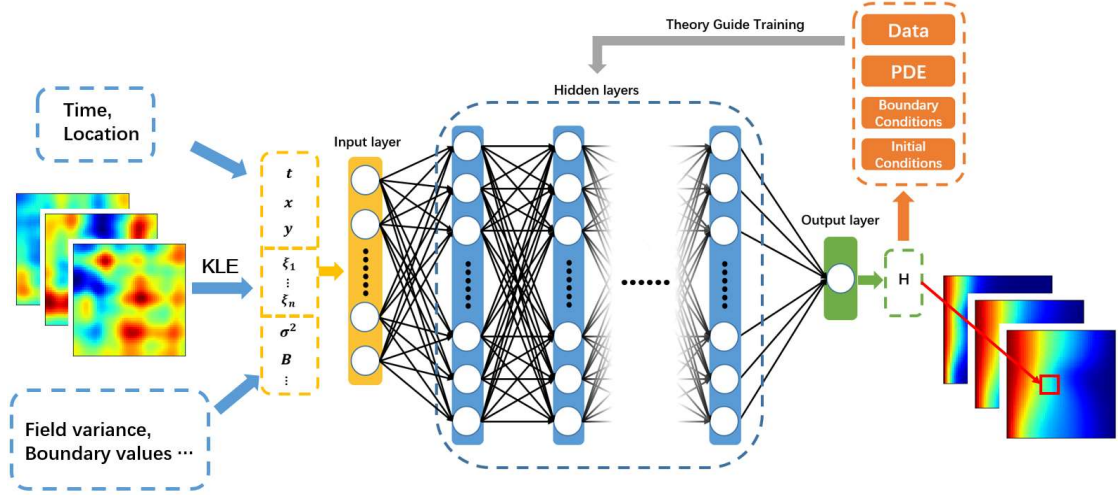

**Fig. 1.** Structure of the TgNN surrogate.

*2.3.3 UQ based on TgNN surrogate*

Once the TgNN surrogate is constructed, UQ tasks can be performed based on the surrogate. In this work, the MC method is adopted to perform UQ via the TgNN surrogate. For any sample of stochastic input, the output can be easily obtained from the TgNN surrogate without the need of solving PDEs. Therefore, the TgNN surrogate can speed up the UQ tasks. The statistical moments and PDFs of the quantity of interest can then be estimated from a certain number of output predictions of the TgNN surrogate.

**3. Cases Studies**

In this section, several subsurface problems are investigated to evaluate the performance of the proposed TgNN surrogate based UQ. The feasibility and accuracy of the TgNN surrogate based UQ is first tested by using a 2D transient saturated flow in porous medium. The effect of the number of collocation points and the labeled-data volume are then studied, so as to assess the worth of labeled data and collocation points. The performance of the TgNN surrogate based UQ on cases with different variances and correlation lengths of the conductivity field is also examined. Finally, to quantify the uncertainty of cases with different boundary values and variances without retraining the surrogate, a composited surrogate is examined.

**3.1 Surrogate for transient saturated subsurface flow in 2D**

A 2D transient saturated flow in porous medium is considered, which satisfies the governing equation of Eq.(1). The domain is a square, with the length in both directions being $L_{x/y} = 1020\,[L]$ (where $[L]$ denotes any consistent length unit). The boundary and initial conditions are as follows:

$$h|_{x=x_0} = 202[L], h|_{x=L_x} = 200[L] \qquad (18)$$

$$\left.\frac{\partial h}{\partial y}\right|_{y=y_0/L_y} = 0 \tag{19}$$

$$h|_{t=0, x \neq x_0} = 200[L] \tag{20}$$

The specific storage is assumed to be a constant, taking a value of $S_s = 0.0001[L^{-1}]$. The mean and variance of the log hydraulic conductivity are given as $\langle \ln K \rangle = 0$ and $\sigma^2_{\ln K} = 1.0$, respectively. Moreover, the correlation length in both directions of the field is $\eta_x = \eta_y = 0.4 L_x = 408[L]$. The hydraulic conductivity field is parameterized through KLE, and approximately 80% energy is maintained, which leads to 20 retained terms in the expansion. Consequently, this field is represented by 20 random variables $\xi = \{\xi_1(\tau), \xi_2(\tau), \cdots, \xi_{20}(\tau)\}$ in this case. Here, MODFLOW software is utilized to numerically solve the problem, in which the domain is evenly divided into 51×51 grid blocks with side length $\Delta x (\Delta y) = 20[L]$, and the total simulation time is $10[T]$ (where $[T]$ denotes any consistent time unit) with each time step being $0.2[T]$, resulting in 50 time steps. Space-time discretization is only used in the numerical model for obtaining the dataset for training; whereas, the TgNN training and prediction are mesh-free, and not restricted to any particular gridding in space or time.

A seven-hidden-layer neural network with 50 neurons per layer is constructed. The time, location and stochastic parameters comprise the input of the neural network, and the hydraulic head is the output. 30 realizations are generated with KLE and solved with MODFLOW, which is represented by $R=30$. 40,000 labeled data points are extracted from each realization to constitute the labeled training dataset, i.e., $N=1.2 \times 10^6$. Furthermore, $10^6$ collocation points are randomly selected in high-dimensional space, as shown below:

$$\begin{aligned} t_i &\sim U(t_0, t_{end}) \\ x_i &\sim U(x_0, x_0 + L_x) \\ y_i &\sim U(y_0, y_0 + L_y) \\ \xi_{1i} &\sim N(0,1) \\ &\vdots \\ \xi_{20i} &\sim N(0,1) \end{aligned} \tag{21}$$

where $U$ and $N$ denote uniform distribution and normal distribution, respectively. The boundary and initial conditions can be incorporated as shown in Eq.(14), Eq.(15), and Eq.(16). While in this case, since the Dirichlet boundary values are constant on each end, a 'hard way' to impose the conditions is adopted, as shown below:

$$\hat{h}(t,x,y,\xi) = h_0\left(1-\frac{x-x_0}{L_x}\right) + h_L\left(\frac{x-x_0}{L_x}\right) + \left(1-\frac{x-x_0}{L_x}\right)\left(\frac{x-x_0}{L_x}\right)NN(t,x,y,\xi;\,\theta)$$

(22)

where $h_0 = 202[L]$ and $h_L = 200[L]$. Here, it is worth noting that a hard way to impose the Dirichlet boundary conditions is convenient in this case, however it may not be suitable for all cases, e.g., cases with varying boundary values. When a hard way for imposing boundary condition is not feasible, the 'soft way' can be applied. In this case, the weights of each term in the loss function are set to be 1, except if otherwise stated. The optimization method Adam (Kingma & Ba, 2015) is utilized with a constant learning rate of 0.001 for 2,000 training epochs to optimize the loss function. The network training is performed on an NVIDIA GeForce RTX 2080 Ti Graphics Processing Unit (GPU) card.

### *3.1.1 Surrogate-based prediction and UQ*

Here, the accuracy of the prediction from the TgNN surrogate is first tested. **Fig. 2** presents the comparison of the hydraulic head distribution at time step 25 predicted via the TgNN surrogate and solved from the MODFLOW simulator for three randomly selected testing realizations for the base case. It can be seen that the TgNN surrogate can provide satisfactory predictions for different hydraulic conductivity fields. **Fig. 3** shows the correlation between the predictions from the surrogate and the reference for two chosen points of 200 testing realizations. It is demonstrated that the predictions from the surrogate can match the reference values well with the plot approaching a line with a 45° angle.

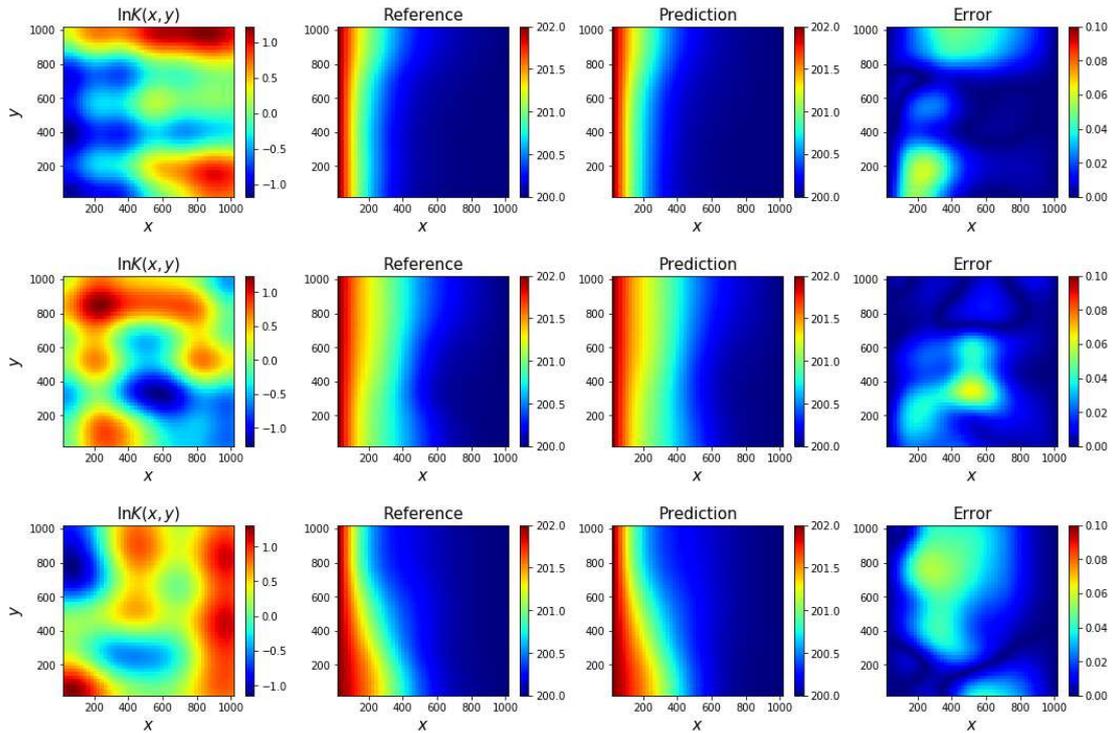

**Fig. 2.** Reference and prediction from the TgNN surrogate at time step 25 for three different hydraulic conductivity fields for the base case.

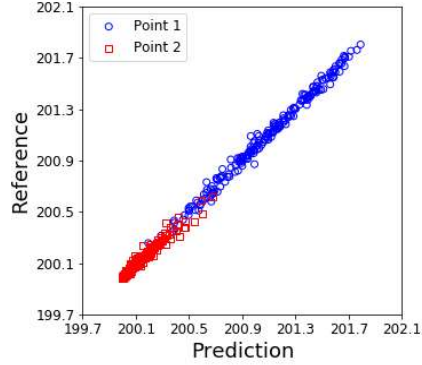

**Fig. 3.** Correlation between predictions from the surrogate and the reference for two points of 200 testing realizations (Point 1: x=200[L], y=200[L], t=5[T]; Point 2: x=800[L], y=200[L], t=8[T]) for the base case.

Furthermore, in order to quantitatively evaluate the results, the following two metrics are introduced. The first is the relative $L_2$ error:

$$L_2(h_{pred}, h_{ref}) = \frac{\|h_{pred} - h_{ref}\|_2}{\|h_{ref}\|_2} \qquad (23)$$

where $h_{ref}$ and $h_{pred}$ denote the reference value solved by MODFLOW and the prediction from the surrogate, respectively; and $\|\cdot\|_2$ denotes the standard Euclidean norm. The second metric is the coefficient of determination, also known as $R^2$ score, which is defined as follows:

$$R^2 = 1 - \frac{\sum_{n=1}^{N_{cell}}(h_{pred,n} - h_{ref,n})^2}{\sum_{n=1}^{N_{cell}}(h_{ref,n} - \overline{h}_{ref})^2} \qquad (24)$$

where $N_{cell}$ denotes the number of blocks needed to be predicted; $h_{ref,n}$ and $h_{pred,n}$ denote the reference value solved by MODFLOW and the prediction from the surrogate at the $n^{th}$ block, respectively; and $\overline{h}_{ref}$ denotes the mean of $h_{ref,n}$. **Fig. 4** presents the histogram of the two metrics for 200 testing realizations, which shows that the TgNN surrogate achieves satisfactory accuracy.

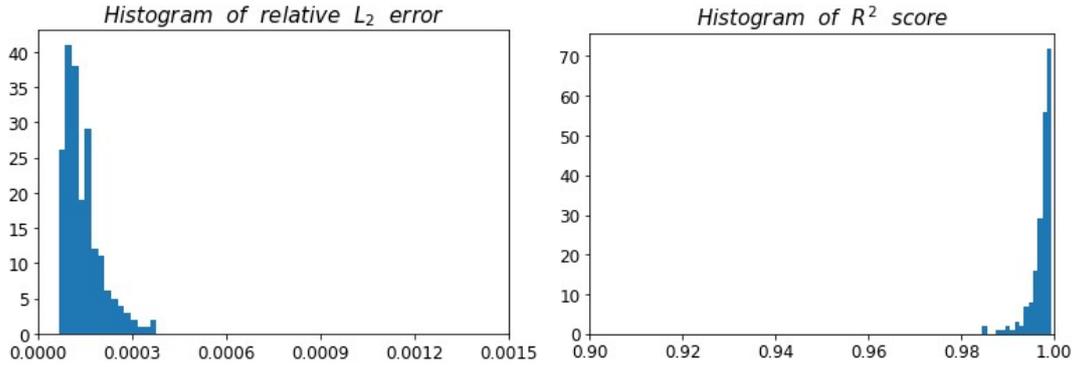

**Fig. 4.** Histograms of relative $L_2$ error and $R^2$ score
for 200 testing realizations for the base case.

The UQ tasks can then be performed based on the constructed TgNN surrogate to easily evaluate statistical quantities, such as mean, variance, and the Probability Density Function (PDF) of the hydraulic head. The MC method is utilized to obtain the benchmark solutions, and 10,000 realizations are generated using KLE with 96% field energy preserved and solved with MODFLOW. The TgNN surrogate is also used to predict the system response for the 10,000 input realizations to estimate the mean, variance, and PDFs of the outputs. The estimated mean and variance from the TgNN surrogate and the MC benchmark at time step 30 are shown in **Fig. 5**. The estimated PDFs of two chosen points are demonstrated in **Fig. 6**. The results show that the TgNN surrogate can accurately quantify the uncertainty, which may also indicate that the trained network has accurately captured the relationship between the input stochastic parameters and the outputs. In addition to accuracy, efficiency is another important criterion for UQ methods. For the investigated case, performing 10,000 simulations with MODFLOW takes approximately 1.75 h (6,304.30 s), while it just takes approximately 9 min (526.14 s) to perform forward calculation 10,000 times with the TgNN surrogate. It can be seen that, once trained, the TgNN surrogate can perform UQ tasks efficiently. It is worth noting that the computational efficiency of the TgNN surrogate may be even more prominent when the size of the model is larger, and that the time to train the TgNN surrogate strongly depends on the number of collocation points and labeled data, which will be discussed in the next subsection.

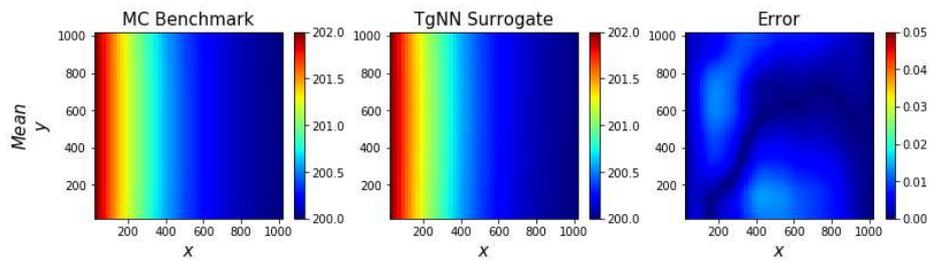

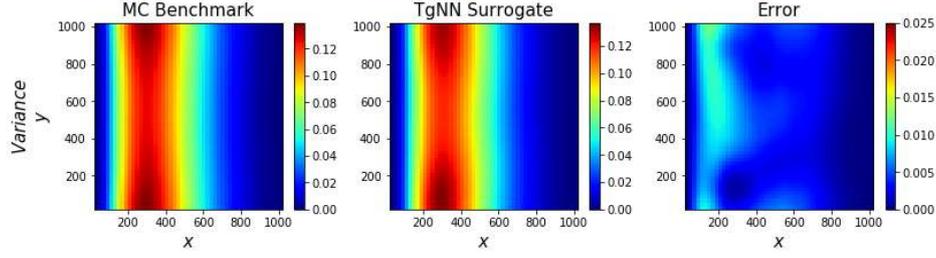

**Fig. 5.** Estimated mean and variance at time step 30 from the TgNN surrogate and the MC benchmark for the base case.

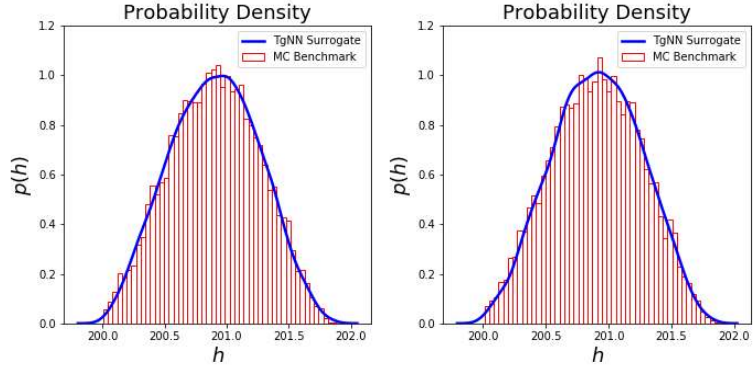

**Fig. 6.** Estimated PDFs for two points (Point 1: x=260[L], y=260[L], t=5[T]; Point 2: x=260[L], y=780[L], t=5[T]) for the base case.

*3.1.2 Effect of collocation points and labeled data*

In order to investigate the performance of the TgNN surrogate more specifically, the effect of the number of collocation points and labeled data is examined. First, the TgNN surrogate is constructed with different numbers of training collocation points ($N_c$) and the same number of labeled data ($R$=30, $N = 1.2 \times 10^6$). **Table 1** shows the relative $L_2$ error and $R^2$ score of the estimated mean and variance of 10,000 realizations from the TgNN surrogate with different settings. In addition, the training time for each TgNN surrogate is also demonstrated in **Table 1**. **Fig. 7** shows the estimated variance from different TgNN surrogates compared with the MC benchmark at time step 30. It can be seen that the TgNN surrogate can provide more accurate UQ results as the number of collocation points increases. However, the time for training a TgNN surrogate increases rapidly with the number of collocation points increasing under a given set of labeled training data. As a consequence, the trade-off between accuracy and computational cost should be taken into account when constructing a TgNN surrogate. For example, while the accuracy is acceptable for the case of $N_c = 1 \times 10^5$ or $5 \times 10^5$, the training time plus the forward UQ calculation time would be much less than the abovementioned time for performing 10,000 MODFLOW runs.

**Table 1.** Relative $L_2$ error and $R^2$ score of estimated mean and variance of 10,000 realizations for different numbers of collocation points with the same number of labeled data ($R=30$, $N = 1.2 \times 10^6$) for the base case.

|  | Mean | | Variance | | Training time (s) |
| --- | --- | --- | --- | --- | --- |
|  | relative $L_2$ error | $R^2$ score | relative $L_2$ error | $R^2$ score |  |
| $N_c = 1 \times 10^5$ | 7.5044e-05 | 9.9931e-01 | 1.7396e-01 | 9.3721e-01 | 1,068.902 |
| $N_c = 5 \times 10^5$ | 7.1267e-05 | 9.9938e-01 | 9.2930e-02 | 9.8208e-01 | 3,995.965 |
| $N_c = 1 \times 10^6$ | 5.6234e-05 | 9.9961e-01 | 7.4768e-02 | 9.8840e-01 | 7,857.762 |

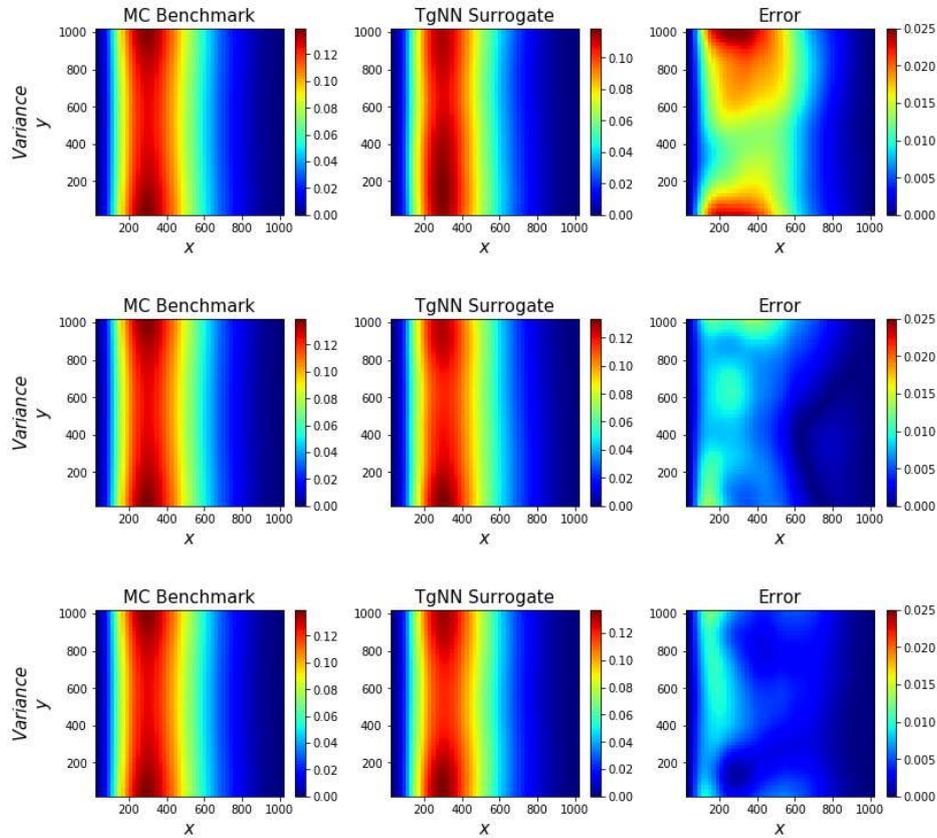

**Fig. 7.** Estimated variance at time step 30 from different TgNN surrogates compared with the MC benchmark for the base case. The TgNN surrogates are trained with $1 \times 10^5$ (first row), $5 \times 10^5$ (second row), and $1 \times 10^6$ (third row) collocation points, respectively.

The TgNN surrogate is then constructed with different numbers of labeled data that are extracted from 0, 10 and 30 realizations, respectively ($R=0, 10, 30, N=0, 4.0 \times 10^6, 1.2 \times 10^6$), and the same number of training collocation points ($N_c = 1 \times 10^6$). The relative $L_2$ error and $R^2$ score of estimated mean and variance from 10,000 realizations are presented in **Table 2**. **Fig. 8** shows a comparison of the estimated variance at time step 30 from different TgNN surrogates with the MC benchmark. It can be seen that the trained TgNN surrogates become more accurate as the number of training labeled data increases. Furthermore, the training time increases slightly with the increase of the number of training data for a given set of collocation points. It is worth noting that the TgNN surrogate can maintain satisfactory accuracy with sparse labeled data (only one realization and all data being extracted), or even with no labeled data ($R=0$, $N=0$), i.e., via label-free surrogate, as discussed in Wang et al. (2020).

Here, only the estimated statistical moments at time step 30 are presented and used for accuracy comparison. To show the capability of the TgNN surrogate for UQ of dynamic subsurface flow, the relative $L_2$ error and $R^2$ score of estimated mean and variance at different time steps are shown in **Appendix B.**

**Table 2.** Relative $L_2$ error and $R^2$ score of estimated mean and variance of 10,000 realizations for different numbers of labeled data with the same number of collocation points ($N_c = 1 \times 10^6$) for the base case.

|  | Mean | | Variance | | Training time (s) |
| --- | --- | --- | --- | --- | --- |
|  | relative $L_2$ error | $R^2$ score | relative $L_2$ error | $R^2$ score |  |
| $R = 0$, $N = 0$ | 1.7752e-04 | 9.9615e-01 | 1.4857e-01 | 9.5420e-01 | 7,585.358 |
| $R = 1$, $N = 130050^*$ | 1.0683e-04 | 9.9861e-01 | 1.2455e-01 | 9.6781e-01 | 7,674.285 |
| $R = 10$, $N = 4 \times 10^5$ | 6.0407e-05 | 9.9955e-01 | 1.0198e-01 | 9.7842e-01 | 7,744.177 |
| $R = 30$, $N = 1.2 \times 10^6$ | 5.6234e-05 | 9.9961e-01 | 7.4768e-02 | 9.8840e-01 | 7,857.762 |

*Note: All of the data in each realization are extracted in this case.

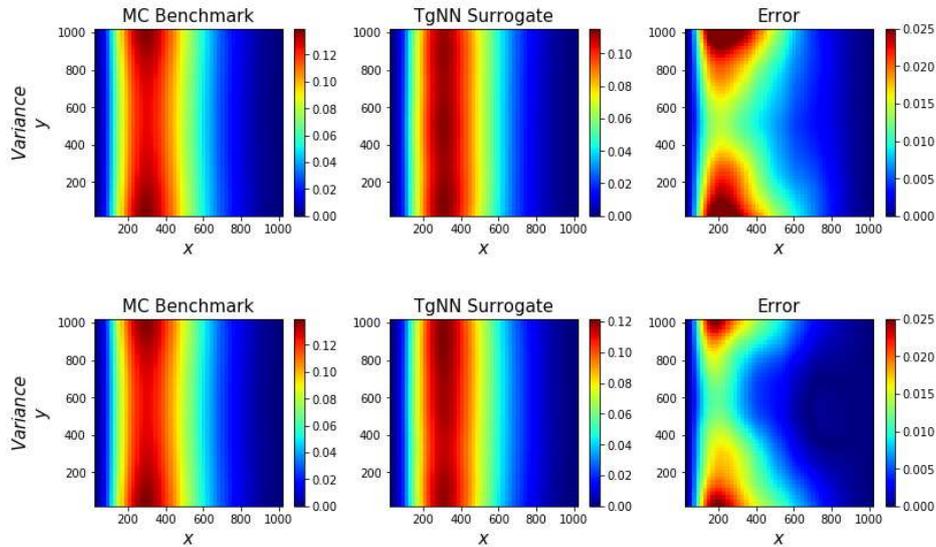

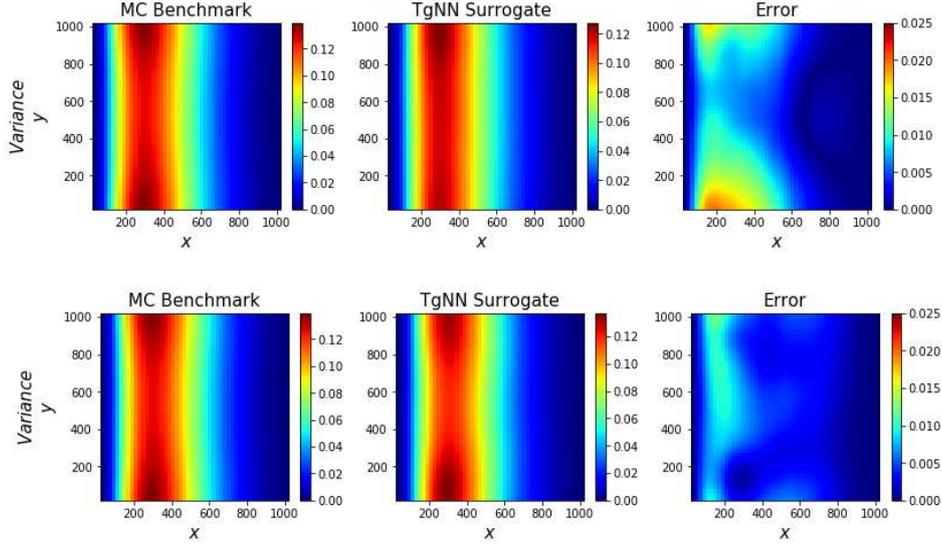

**Fig. 8.** Estimated variance at time step 30 from different TgNN surrogates compared with the MC benchmark for the base case. The TgNN surrogates are trained with 0 (first row), 130,050 (second row), $4 \times 10^5$ (third row), and $1.2 \times 10^6$ (fourth row) labeled data, respectively.

### *3.1.3 Label-free TgNN surrogate*

It has been demonstrated above that the TgNN surrogate still works even in label-free scenarios. Here this extreme situation is investigated further. The TgNN surrogates are constructed with different numbers of collocation points and no labeled data from the simulation. The weight of the PDE loss term in the loss function is increased to 10 to enforce stronger physical constraints. **Appendix A** shows the estimated mean and variance for 10,000 realizations with label-free TgNN surrogates. **Table 3** presents the relative $L_2$ errors and $R^2$ scores.

**Table 3.** Relative $L_2$ error and $R^2$ score of estimated mean and variance of 10,000 realizations for different numbers of collocation points in the case of label-free training.

|  | **Mean** |  | **Variance** |  | **Training time (s)** |
|---|---|---|---|---|---|
|  | relative $L_2$ error | $R^2$ score | relative $L_2$ error | $R^2$ score |  |
| $N_c = 1 \times 10^4$ | 1.4129e-04 | 9.9756e-01 | 3.5733e-01 | 7.3506e-01 | 374.636 |
| $N_c = 5 \times 10^5$ | 1.8922e-04 | 9.9563e-01 | 1.2532e-01 | 9.6741e-01 | 3,973.966 |
| $N_c = 1 \times 10^6$ | 1.1798e-04 | 9.9830e-01 | 1.0416e-01 | 9.7749e-01 | 7,651.414 |
| $N_c = 1.5 \times 10^6$ | 8.0967e-05 | 9.9920e-01 | 7.8432e-02 | 9.8724e-01 | 11,230.331 |

It is obvious that the TgNN surrogates can provide accurate UQ results, relying solely on physical constraints given that an adequate number of collocation points are employed. Therefore, the TgNN surrogate is applicable even under circumstances in which no simulation data are available. Comparing the results shown in **Table 1, Table 2** and **Table 3**, it can be seen that the computation cost increases much more rapidly with the number of collocation points increasing than that with the number of labeled data increasing. For example, for the case of $N_c = 1 \times 10^6$ shown in **Table 3**, the training time is similar to the last two cases shown in **Table 2**, but the accuracy is lower; for the case of $N_c = 1.5 \times 10^6$, the accuracy is similar to the abovementioned two cases shown in **Table 2**, but the training time is much larger. Consequently, when quality labeled data are available, it is better to incorporate them together with physical constraints in the training process to construct a surrogate. Apparently, incorporating as much available information as possible to construct a surrogate is advantageous, and the benefit of training data will be more significant when forward simulations are costly.

**3.2 Increasing heterogeneity of the stochastic field**

To further investigate the robustness of the TgNN surrogate based UQ, the heterogeneity of the stochastic field is increased, and cases with shorter correlation lengths and higher variances, respectively, are considered in this subsection.

*3.2.1 Decreasing the correlation length*

The effect of correlation length is first studied here. The correlation length of the case in subsection 3.1 is reduced to $\eta_x = \eta_y = 0.2 L_x = 204 [L]$, and 71 terms are retained in KLE to preserve 80% field energy. As a result, together with time and location, there are 74 terms in the input. A seven-hidden-layer neural network with 100 neurons per layer is constructed for this case. 80 realizations are generated with KLE and solved with MODFLOW to provide training labeled data, and 40,000 labeled data points are extracted from each realization. In addition, $3 \times 10^6$ collocation points are randomly selected in high-dimensional space for training. The training process for this case takes approximately 4.6 h (16,571.87 s) on the NVIDIA GeForce RTX 2080 Ti GPU.

**Fig. 9** shows the predicted hydraulic distribution at time step 25 of three randomly sampled realizations and the reference solved from MODFLOW. **Fig. 10** presents the histogram of the relative $L_2$ errors and $R^2$ scores of 200 testing realizations. It can be seen that $R^2$ scores approach to 1, and relative $L_2$ errors are close to 0. **Fig. 11** shows the statistical moments and PDFs of three points estimated from 10,000 realizations. For the MC benchmark, 95% energy is preserved to guarantee accuracy. The results demonstrate that the TgNN surrogate can provide accurate prediction and statistical quantity estimation for cases with shorter correlation lengths. It is worth mentioning that the case considered has large random dimensionality, *n*=71. Large dimensionality constitutes a major challenge for most UQ methods, such as Polynomial Chaos Expansion (PCE). For second, third, and forth order PCE with random dimensionality being 71, the required simulations are 2,628, 64,824, and 1,215,450, respectively. It is clear that, for high-dimensional problems, high order PCE is not affordable regarding computational cost. However, the TgNN surrogate shows promising applicability for UQ of high-dimensional problems.

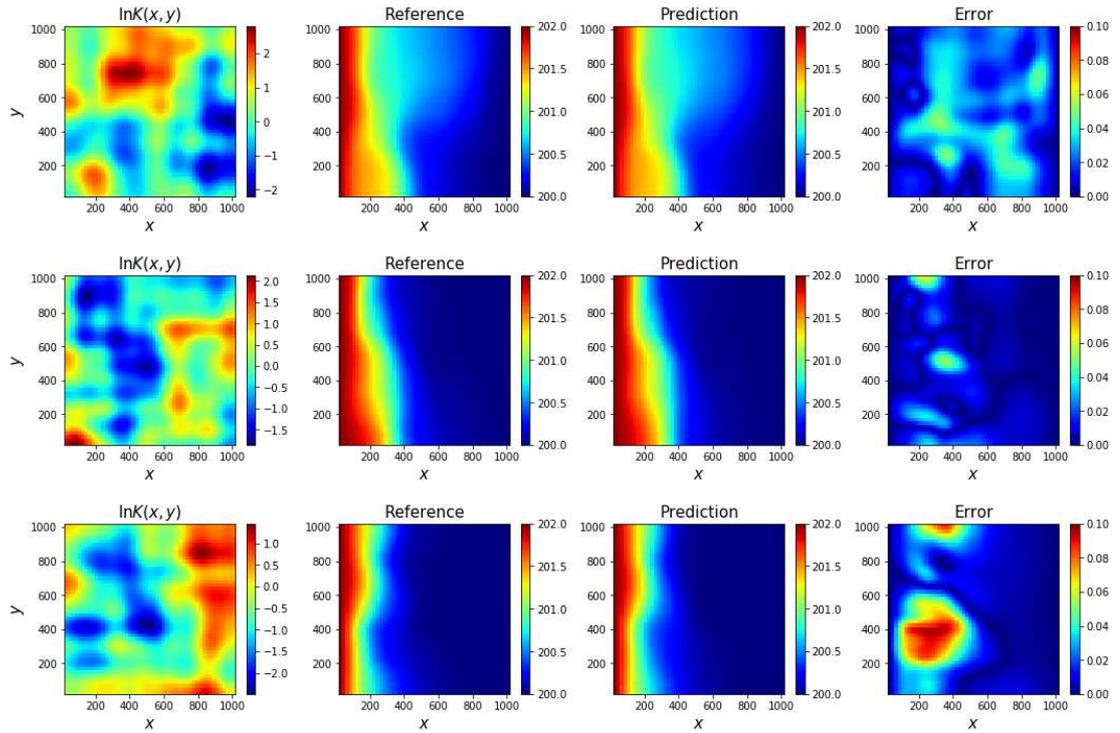

**Fig. 9.** Reference and prediction from the TgNN surrogate at time step 25 for three different hydraulic conductivity fields for the case of $\eta_x = \eta_y = 0.2L_x = 204\,[L]$.

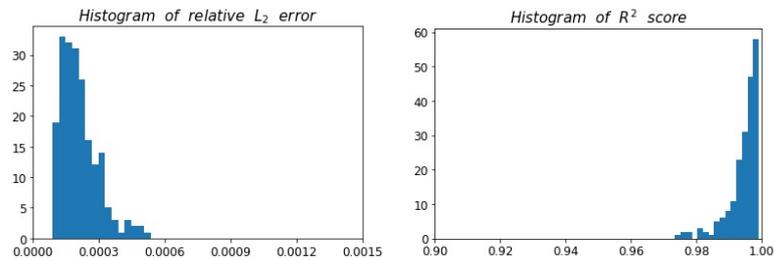

**Fig. 10.** Histograms of relative $L_2$ error and $R^2$ score for 200 testing realizations for the case of $\eta_x = \eta_y = 0.2L_x = 204\,[L]$.

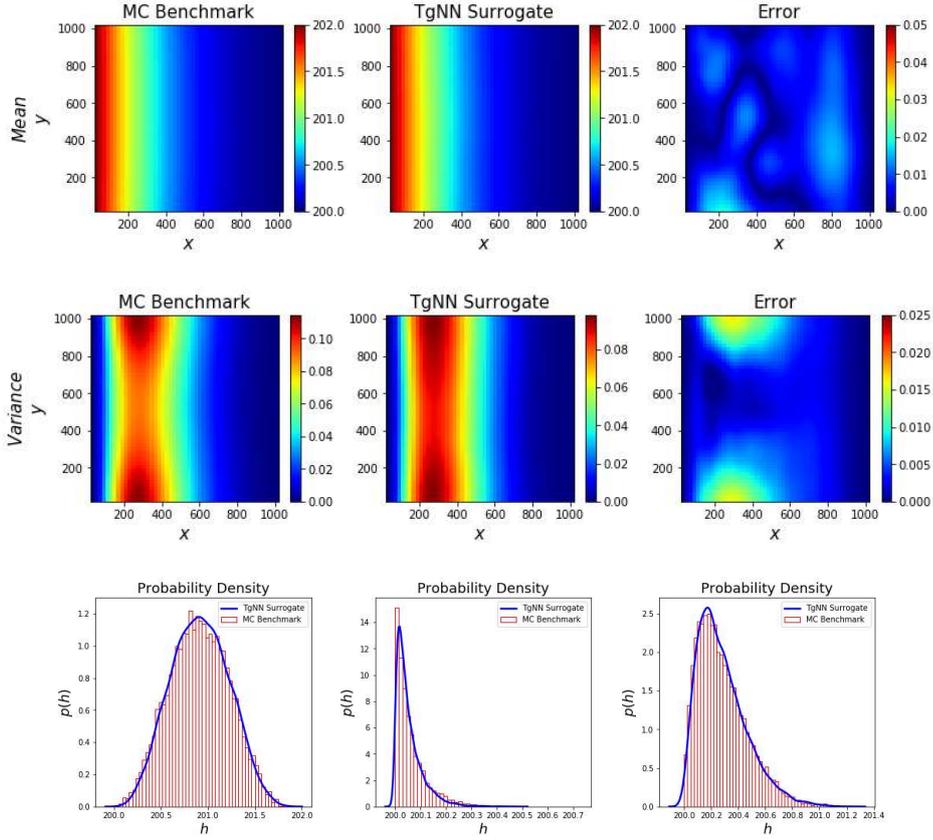

**Fig. 11.** Estimated statistical moments at time step 30 and PDFs for three points (Point 1: x=260[L], y=260[L], t=5[T]; Point 2: x=780 [L], y=260 [L], t=5[T]; Point 3: x=520[L], y=520[L], t=5[T]) from the TgNN surrogate and the MC benchmark for the case of $\eta_x = \eta_y = 0.2L_x = 204[L]$.

### 3.2.2 Increasing the variance

The variance of the stochastic field is increased to $\sigma^2_{\ln K} = 2.0$ in this case. There are still 20 terms retained in KLE, and 80 neurons per layer in a seven-hidden-layer neural network. 30 realizations are solved with the simulator to provide training data, and $2 \times 10^6$ collocation points are randomly extracted in high-dimensional space for training. It takes approximately 5.6 h (20,131.12 s) to train the neural network on the NVIDIA GeForce RTX 2080 Ti GPU.

**Fig. 12** shows the predicted hydraulic distribution at time step 25 of three randomly sampled realizations and the reference solved from MODFLOW. **Fig. 13** presents the histogram of the relative $L_2$ errors and $R^2$ scores of 200 testing realizations. **Fig. 14** shows the estimated statistical moments of hydraulic heads from 10,000 realizations. It can be seen that the TgNN surrogate can provide accurate prediction and statistical quantity estimation for cases with higher variances.

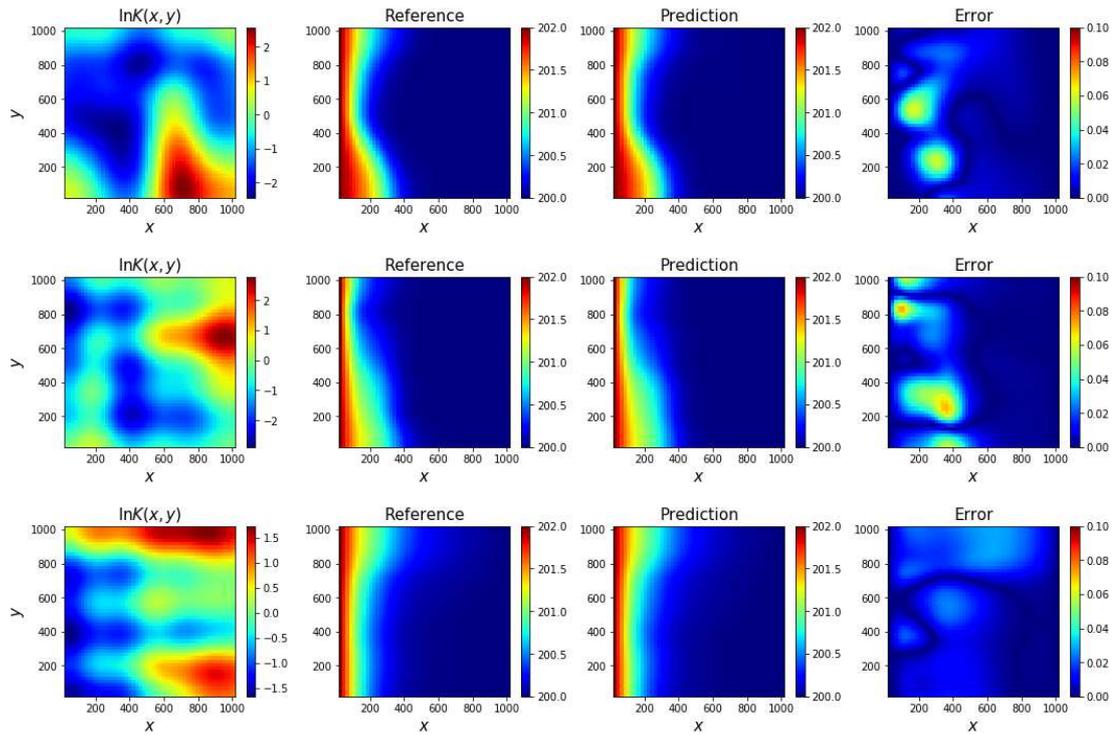

**Fig. 12.** Reference and prediction from the TgNN surrogate at time step 25 for different hydraulic conductivity fields for the case of $\sigma^2_{\ln K} = 2.0$.

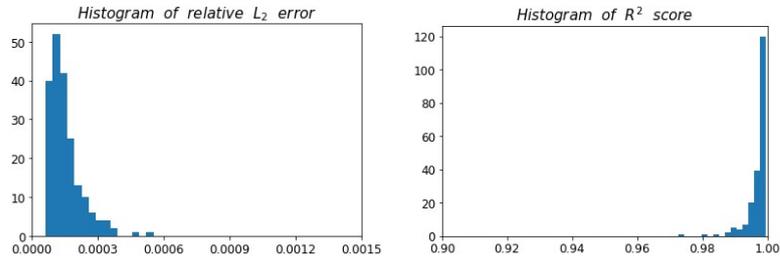

**Fig. 13.** Histograms of relative $L_2$ error and $R^2$ score for 200 testing realizations for the case of $\sigma^2_{\ln K} = 2.0$.

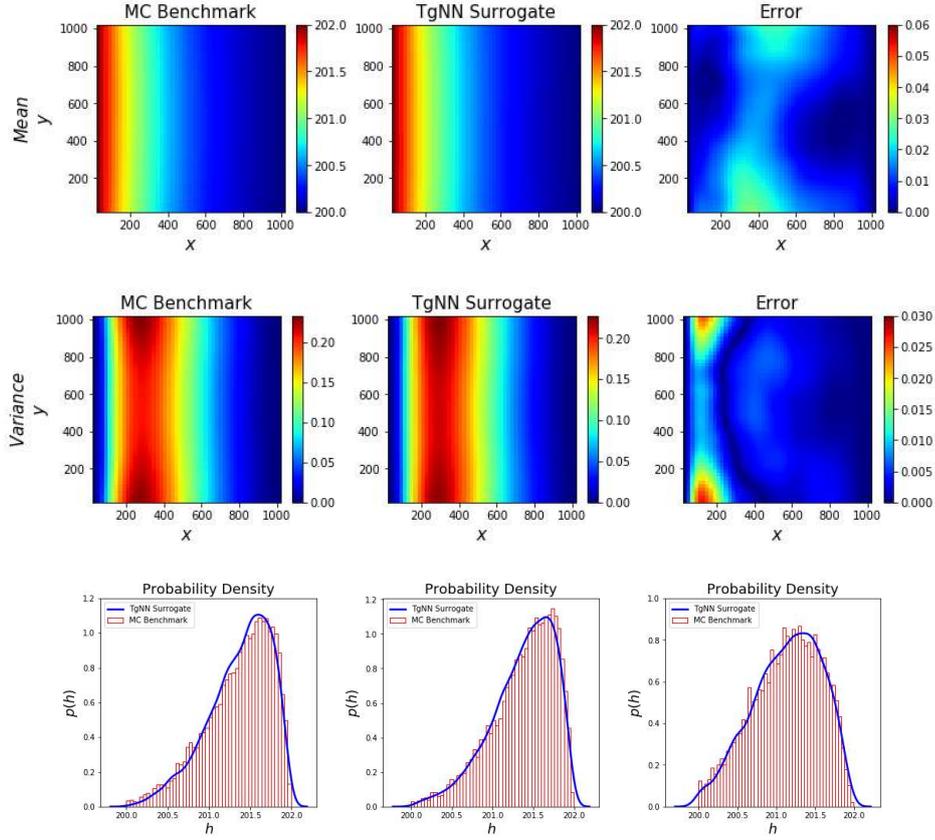

**Fig. 14.** Estimated statistical moments at time step 30 and PDFs for three points (Point 1: x=140[L], y=140[L], t=5.2[T]; Point 2: x=140 [L], y=780 [L], t=5.2[T]; Point 3: x=260[L], y=780[L], t=9.2[T]) from the TgNN surrogate and the MC benchmark for the case of $\sigma^2_{\ln K} = 2.0$.

### 3.3 A composite surrogate with changing boundary values and variances

In the former cases, the surrogates are constructed with specific boundary values and stochastic field variances. Every time the boundary value or the variance is changed, however, the surrogate should be retrained. In order to alleviate this problem, a composite surrogate is proposed in this section, which incorporates the boundary value and variance as inputs to the neural network. Consequently, the trained surrogate can be used to make predictions for cases with different boundary values and field variances directly.

In this case, the variance of the stochastic filed is supposed to follow a uniform distribution, $\sigma^2_{\ln K} \sim U(1, 2)$. The prescribed hydraulic heads of the two boundaries ($x = x_0, x = L_x$) are assumed to follow a normal distribution $h_0 \sim N(202, 0.25)[L]$ and

$h_L \sim N(200, 0.25)[L]$, respectively. The collocation points $\{t_i, \mathbf{x}_i, \boldsymbol{\xi}_i, B_{1i}, B_{2i}, \sigma_i^2\}_{i=1}^{N_c}$ are randomly extracted in high-dimensional parameter space, where $B_1, B_2$ and $\sigma^2$ denote the two prescribed hydraulic heads on the two boundaries and the variance, respectively. A seven-hidden-layer network is constructed with 80 neurons in each layer. 150 realizations with different boundary values and field variances are solved with MODFLOW to provide $4 \times 10^6$ labeled data with 40,000 in each realization. The number of collocation points is $N_c = 2 \times 10^6$. The training process of the composited surrogate takes approximately 6.1 h (21,825.39 s) on an NVIDIA GeForce RTX 2080 Ti GPU.

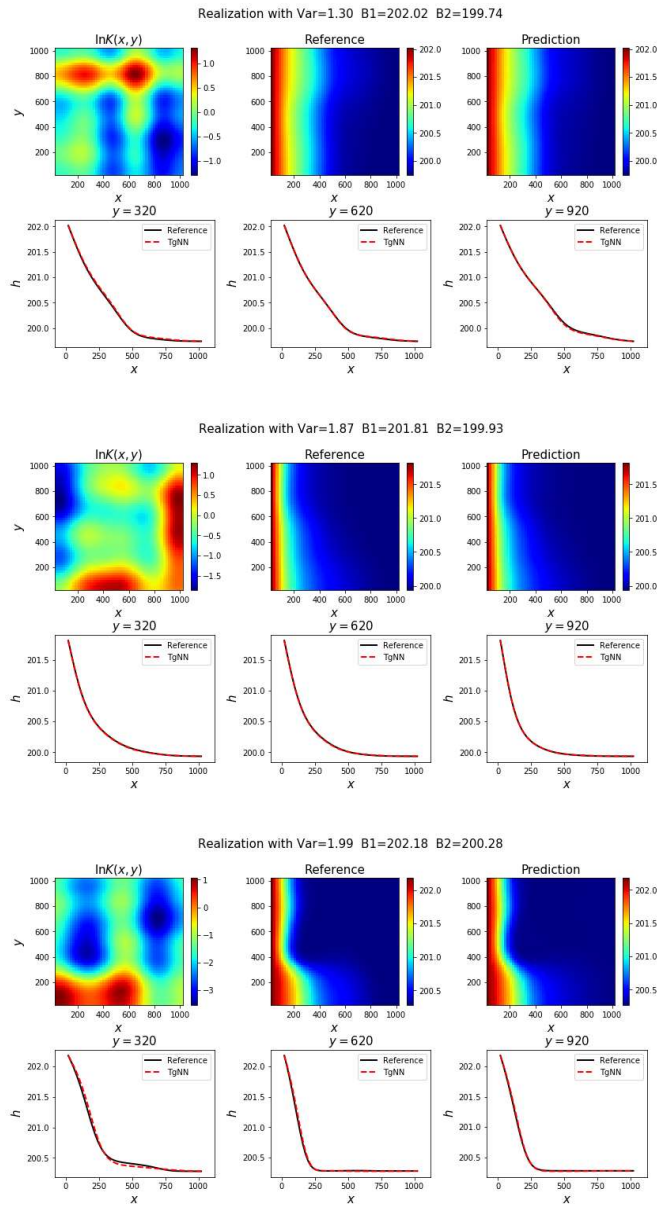

**Fig. 15.** Reference and prediction from the TgNN surrogate at time step 25 for three different sampled realizations for the case of composite surrogate.

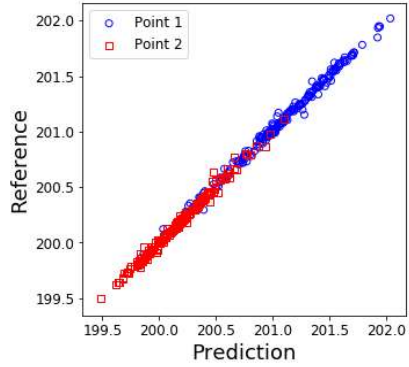

**Fig. 16.** Correlation between the predictions from the TgNN surrogate and the reference for two points of 200 testing realizations (Point 1: x=200[L], y=200[L], t=5[T]; Point 2: x=800[L], y=200[L], t=8[T]) for the case of composite surrogate.

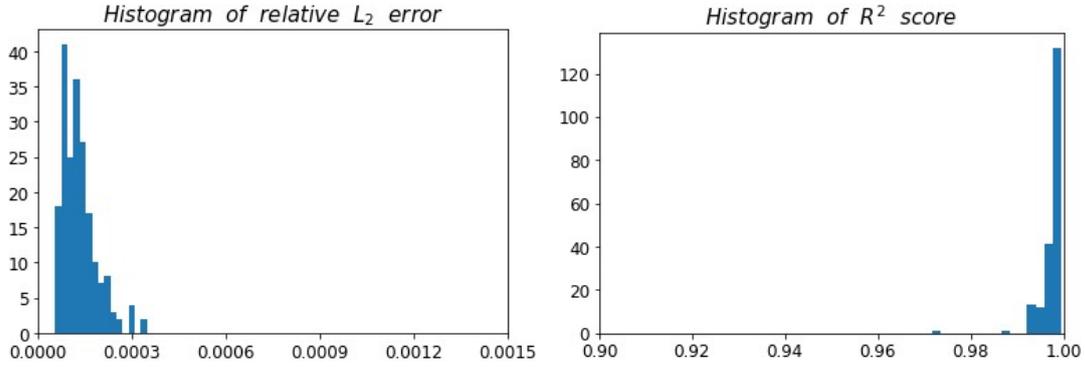

**Fig. 17.** Histograms of relative $L_2$ error and $R^2$ score for 200 testing realizations for the case of composite surrogate.

The trained composite surrogate can make predictions for cases with different boundary values and field variances, as well as different realizations of hydraulic conductivity (defined by Eq. (7)). **Fig. 15** shows the predictions at time step 25 of three sampled realizations with randomly selected parameters in high-dimensional space. **Fig. 16** presents the correlation of reference and predictions for two chosen points of 200 different sampled realizations of varying hydraulic conductivity fields, boundary values, and field variances. **Fig. 17** presents the histogram of the relative $L_2$ errors and $R^2$ scores of those testing realizations.

*3.3.1 UQ for cases with different variances*

Given the composite surrogate, UQ tasks can be performed at a minimum cost for different boundary values or field variances. Compared to needing to make a large number of new forward simulations with a simulator in the traditional Monte Carlo approach for a new boundary value or field variance, sampling with the composite surrogate is extremely fast. For example, one can fix the variance and boundary values in the input and change the other parameters to achieve UQ tasks. The estimated statistical moments and PDFs from 10,000 samples with variance = 1 and variance = 2 are shown in **Fig. 18** and **Fig. 19**, respectively. Compared with the results shown in subsection 3.1.1 and 3.2.2, the composite surrogate makes satisfactory estimations with higher efficiency.

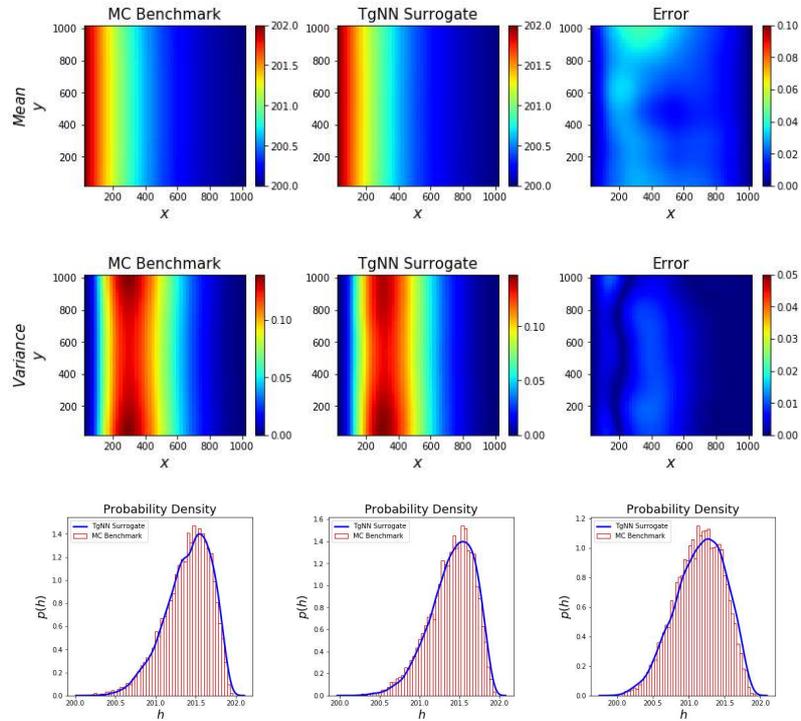

**Fig. 18.** Estimated statistical moments at time step 30 and PDFs for three points (Point 1: x=140[L], y=140[L], t=5.2[T]; Point 2: x=140 [L], y=780 [L], t=5.2[T]; Point 3: x=260[L], y=780[L], t=9.2[T]) from the composite TgNN surrogate and the MC benchmark when variance = 1.

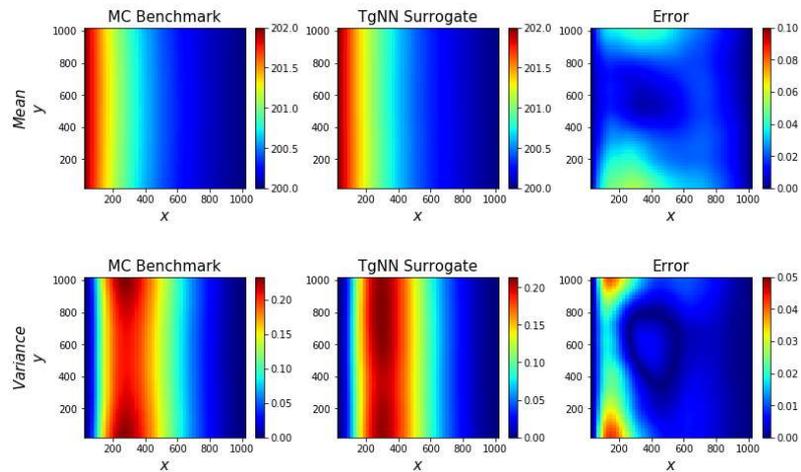

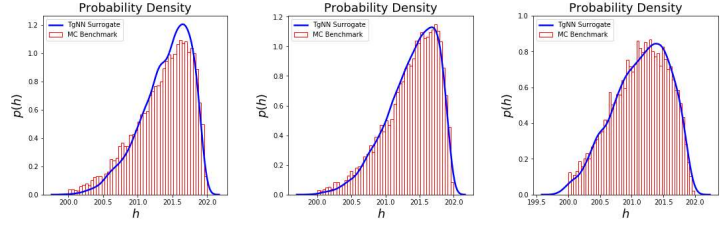

**Fig. 19.** Estimated statistical moments at time step 30 and PDFs for three points (Point 1: x=140[L], y=140[L], t=5.2[T]; Point 2: x=140 [L], y=780 [L], t=5.2[T]; Point 3: x=260[L], y=780[L], t=9.2[T]) from the composite TgNN surrogate and the MC benchmark when variance = 2.

*3.3.2 Extrapolation and transfer learning for out-of-distribution variances*

The collocation points are sampled in the interval [1, 2] for variance to train the composite surrogate as aforementioned. Although not shown, the composite TgNN surrogate possesses a strong capability to interpolate within this interval from which the training data and collocation points came. Here, in order to investigate the extrapolation performance of the composite surrogate, cases with variances out-of-distribution are studied. For each specific variance, 200 realizations are randomly sampled to calculate the mean and variance of the outputs using the trained composite surrogate. **Table 4** presents the relative $L_2$ error and $R^2$ score of statistical moments for 200 sampled realizations with out-of-distribution field variances. It can be seen that the trained composite surrogate has limited capacity of extrapolation for variances out-of-distribution. In addition, the extrapolation performance becomes worse as the variance gets further away from the preset interval.

In order to improve the extrapolation performance, a transfer learning strategy (Pan & Yang, 2009) is adopted here for cases in which the variance is too far away from its preset interval. The transfer learning strategy has been introduced and studied for TgNN when dealing with changing boundary situations (Wang et al., 2020). In the transfer learning process, the pre-trained composite TgNN surrogate is used for initialization, and a few new collocation points are sampled for fine-tuning the pre-trained network. The new collocation points are sampled under the target variance, and no labeled data are needed in this training process.

**Table 4.** Relative $L_2$ error and $R^2$ score of calculated statistical moments for 200 realizations with different field variances by the composite TgNN surrogate.

|  | Mean | | Variance | |
|---|---|---|---|---|
|  | relative $L_2$ error | $R^2$ score | relative $L_2$ error | $R^2$ score |
| $\sigma^2_{\ln K} = 0.4$ | 8.4004e-05 | 9.9915e-01 | 7.3621e-01 | -1.8306e-02 |
| $\sigma^2_{\ln K} = 0.6$ | 7.0272e-05 | 9.9939e-01 | 3.3280e-01 | 7.8828e-01 |
| $\sigma^2_{\ln K} = 0.8$ | 6.1833e-05 | 9.9952e-01 | 1.5006e-01 | 9.5620e-01 |
| $\sigma^2_{\ln K} = 2.2$ | 5.8998e-05 | 9.9952e-01 | 4.9748e-02 | 9.9462e-01 |
| $\sigma^2_{\ln K} = 2.4$ | 6.2488e-05 | 9.9946e-01 | 5.4530e-02 | 9.9345e-01 |
| $\sigma^2_{\ln K} = 2.6$ | 6.8511e-05 | 9.9934e-01 | 6.7935e-02 | 9.8969e-01 |

Take the scenario with $\sigma^2_{\ln K} = 0.4$ as an example. $10^5$ collocation points, i.e., $\{t_i, \mathbf{x}_i, \xi_i, B_{1i}, B_{2i}, \sigma^2_i = 0.4\}_{i=1}^{N_c=10^5}$, are sampled for training, and the network is initialized with the pre-trained model. The network is trained for just 200 epochs. **Table 5** shows the relative $L_2$ error and $R^2$ score of statistical moments for 200 sampled realizations of two cases ($\sigma^2_{\ln K} = 0.4$ and $\sigma^2_{\ln K} = 0.6$). **Fig. 20** presents the correlation of reference and predictions for two points of the 200 sampled realizations. It can be seen that the fine-tuned surrogate can make accurate predictions for cases with out-of-distribution variance and the training process only takes approximately 1.5 min, which is diminutive compared with retraining the network. Therefore, in aid of transfer learning strategy, the composite surrogate can make more accurate predictions for input that is beyond its preset interval where the collocation points are sampled.

**Table 5.** Relative $L_2$ error and $R^2$ score of statistical moments for 200 sampled realizations of the two out-of-distribution variances by transfer learning.

| | Mean | | Variance | | Training time (s) |
|---|---|---|---|---|---|
| | relative $L_2$ error | $R^2$ score | relative $L_2$ error | $R^2$ score | |
| $\sigma^2_{\ln K} = 0.4$ | 6.0603e-05 | 9.9956e-01 | 5.5010e-02 | 9.9431e-01 | 99.23 |
| $\sigma^2_{\ln K} = 0.6$ | 5.9663e-05 | 9.9956e-01 | 4.9899e-02 | 9.9524e-01 | 99.27 |

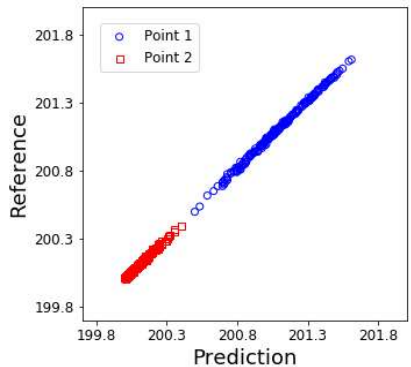 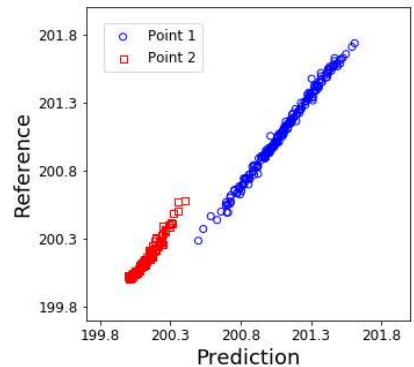

a) Transfer learning for $\sigma^2_{\ln K} = 0.4$    b) Just extrapolation for $\sigma^2_{\ln K} = 0.4$

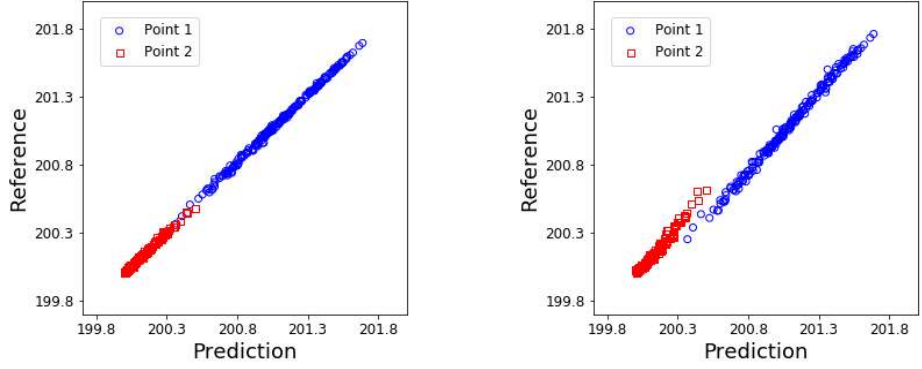

c) Transfer learning for $\sigma^2_{\ln K} = 0.6$    d) Just extrapolation for $\sigma^2_{\ln K} = 0.6$

**Fig. 20.** Correlation between the predictions from the surrogate and the reference for two points (Point 1: x=200[L], y=200[L], t=5[T]; Point 2: x=800[L], y=200[L], t=8[T]) with and without transfer learning.

## 4. Discussions and Conclusions

In this work, we proposed a methodology for efficient uncertainty quantification (UQ) for dynamic subsurface flow with a surrogate constructed by the Theory-guided Neural Network (TgNN). The TgNN is specially designed for problems with stochastic parameters. In addition to time and location, the input of TgNN also includes stochastic parameters. In the training process, physical theories are incorporated into the loss function, such as governing equation, boundary condition, initial condition, etc. The trained TgNN can work as a surrogate for predicting solutions of dynamic subsurface flow problems with new stochastic parameters. UQ tasks can then be efficiently implemented with the assistance of the TgNN surrogate. Transient subsurface flow problems are utilized to test the performance of the TgNN surrogate based UQ, and satisfactory results are obtained.

The number of labeled data and the number of collocation points are important factors for the TgNN surrogate. Numerical results demonstrate that more training data or collocation points can lead to higher accuracy, but the computational cost will increase. In addition, the computation cost increases much more rapidly with the number of collocation points increasing than that with the number of labeled data increasing. A trade-off between efficiency and accuracy should be considered when constructing the TgNN surrogate. On the other hand, the TgNN surrogate can work in a label-free manner. The label-free TgNN can achieve satisfactory UQ results with utilization of adequate collocation points. Even though label-free TgNN works well, when quality labeled data are available, incorporating them together with physical constraints is a superior option.

In this work, a case with high random dimensionality ($n$=71) is investigated, and satisfactory UQ results are obtained from the TgNN surrogate. The potential for solving high-dimensional problems constitutes an advantage of the TgNN surrogate. The input of the proposed TgNN surrogate can include various uncertain parameters. Cases with uncertain boundary values and field variances are also examined. By incorporating the boundary values and field variances into the input, there is no need to retrain the network totally when the boundary values and field variances are changed. Furthermore, the transfer learning strategy can be adopted for cases with out-of-distribution values. The flexibility of considering different uncertain parameters and applying transfer learning is the superior feature of TgNN surrogate based UQ.

**Acknowledgements**

This work is partially funded by the National Natural Science Foundation of China (Grant No. 51520105005 and U1663208) and the National Science and Technology Major Project of China (Grant No. 2017ZX05009-005 and 2017ZX05049-003).

**References**

Awotunde, A. A., & Horne, R. N. (2013). Reservoir description with integrated multiwell data using two-dimensional wavelets. *Mathematical Geosciences, 45*(2), 225-252.

Ballio, F., & Guadagnini, A. (2004). Convergence assessment of numerical Monte Carlo simulations in groundwater hydrology. *Water Resources Research, 40*(4).

Bottou, L. (2010). Large-scale machine learning with stochastic gradient descent. In *Proceedings of COMPSTAT'2010* (pp. 177-186): Springer.

Chang, H., & Zhang, D. (2009). A comparative study of stochastic collocation methods for flow in spatially correlated random fields. *Communications in Computational Physics, 6*(3), 509.

Ghanem, R. G., & Spanos, P. D. (2003). *Stochastic finite elements: a spectral approach*: Courier Corporation.

Goodfellow, I., Bengio, Y., & Courville, A. (2016). *Deep learning*: MIT press.

Jafarpour, B., Goyal, V. K., McLaughlin, D. B., & Freeman, W. T. (2010). Compressed history matching: exploiting transform-domain sparsity for regularization of nonlinear dynamic data integration problems. *Mathematical Geosciences, 42*(1), 1-27.

Karumuri, S., Tripathy, R., Bilionis, I., & Panchal, J. (2020). Simulator-free solution of high-dimensional stochastic elliptic partial differential equations using deep neural networks. *Journal of Computational Physics, 404*, 109120.

Kennedy, M. C., & O'Hagan, A. (2000). Predicting the output from a complex computer code when fast approximations are available. *Biometrika, 87*(1), 1-13.

Kingma, D. P., & Ba, J. L. (2015). Adam: A Method for Stochastic Optimization. Paper presented at the International conference on learning representations.

Li, H., & Zhang, D. (2007). Probabilistic collocation method for flow in porous media: Comparisons with other stochastic methods. *Water Resources Research, 43*(9).

Li, H., & Zhang, D. (2013). Stochastic representation and dimension reduction for non-Gaussian random fields: review and reflection. *Stochastic Environmental Research and Risk Assessment, 27*(7), 1621-1635.

Liao, Q., & Zhang, D. (2015). Constrained probabilistic collocation method for uncertainty quantification of geophysical models. *Computational Geosciences, 19*(2), 311-326.

Mo, S., Zabaras, N., Shi, X., & Wu, J. (2019). Deep autoregressive neural networks for high‐dimensional inverse problems in groundwater contaminant source identification. *Water Resources Research, 55*(5), 3856-3881.

Mo, S., Zhu, Y., Zabaras, N., Shi, X., & Wu, J. (2019). Deep convolutional encoder‐decoder networks for uncertainty quantification of dynamic multiphase flow in heterogeneous media. *Water Resources Research, 55*(1), 703-728.

Pan, S. J., & Yang, Q. (2009). A survey on transfer learning. *IEEE Transactions on Knowledge Data Engineering, 22*(10), 1345-1359.

Park, J., & Sandberg, I. W. (1991). Universal approximation using radial-basis-function networks. *Neural Computation, 3*(2), 246-257.

Raissi, M., Perdikaris, P., & Karniadakis, G. E. (2019). Physics-informed neural networks: A deep learning framework for solving forward and inverse problems involving nonlinear partial differential equations. *Journal of Computational Physics, 378*, 686-707.

Ramachandran, P., Zoph, B., & Le, Q. V. (2017). Searching for activation functions. *arXiv preprint arXiv:.05941*.

Regis, R. G., & Shoemaker, C. A. (2007). A stochastic radial basis function method for the global optimization of expensive functions. *INFORMS Journal on Computing, 19*(4), 497-509.

Smith, R. C. (2013). *Uncertainty quantification: theory, implementation, and applications* (Vol. 12): Siam.

Sun, L., Gao, H., Pan, S., Wang, J.-X., & Engineering. (2020). Surrogate modeling for fluid flows based on physics-constrained deep learning without simulation data. *Computer Methods in Applied Mechanics Engineering, 361*, 112732.

Tavakoli, R., & Reynolds, A. C. (2011). Monte Carlo simulation of permeability fields and reservoir performance predictions with SVD parameterization in RML compared with EnKF. *Computational Geosciences, 15*(1), 99-116.

Tripathy, R. K., & Bilionis, I. (2018). Deep UQ: Learning deep neural network surrogate models for high dimensional uncertainty quantification. *Journal of Computational Physics, 375*, 565-588.

Wang, N., Zhang, D., Chang, H., & Li, H. (2020). Deep learning of subsurface flow via theory-guided neural network. *Journal of Hydrology, 584*, 124700.

Wang, Y., Yao, H., & Zhao, S. (2016). Auto-encoder based dimensionality reduction. *Neurocomputing, 184*, 232-242.

Williams, C. K., & Rasmussen, C. E. (2006). *Gaussian processes for machine learning* (Vol. 2): MIT press Cambridge, MA.

Xiu, D., & Karniadakis, G. E. (2002a). Modeling uncertainty in steady state diffusion problems via generalized polynomial chaos. *Computer Methods in Applied Mechanics Engineering, 191*(43), 4927-4948.

Xiu, D., & Karniadakis, G. E. (2002b). The Wiener--Askey polynomial chaos for stochastic differential equations. *SIAM Journal on Scientific Computing, 24*(2),

619-644.

Zhang, D. (2001). *Stochastic methods for flow in porous media: coping with uncertainties*: Elsevier.

Zhang, D., & Lu, Z. (2004). An efficient, high-order perturbation approach for flow in random porous media via Karhunen–Loeve and polynomial expansions. *Journal of Computational Physics, 194*(2), 773-794.

Zhu, Y., & Zabaras, N. (2018). Bayesian deep convolutional encoder–decoder networks for surrogate modeling and uncertainty quantification. *Journal of Computational Physics, 366*, 415-447.

Zhu, Y., Zabaras, N., Koutsourelakis, P.-S., & Perdikaris, P. (2019). Physics-constrained deep learning for high-dimensional surrogate modeling and uncertainty quantification without labeled data. *Journal of Computational Physics, 394*, 56-81.

**Appendix A**

In this appendix, the estimated statistical moments from the label-free TgNN surrogate are provided. In the considered case, no labeled data are needed, and the physical theories (PDEs, initial and boundary conditions) are solely considered to train the TgNN surrogate. The label-free TgNN surrogate is trained with different numbers of collocation points ($1 \times 10^4$, $5 \times 10^5$, $1 \times 10^6$, and $1.5 \times 10^6$), and the estimated statistical moments of time step 30 from 10,000 realizations are shown in **Fig. A.1** and **Fig. A.2**. It can be seen that the label-free TgNN surrogate can provide accurate UQ results, and accuracy will increase as the number of collocation points increases.

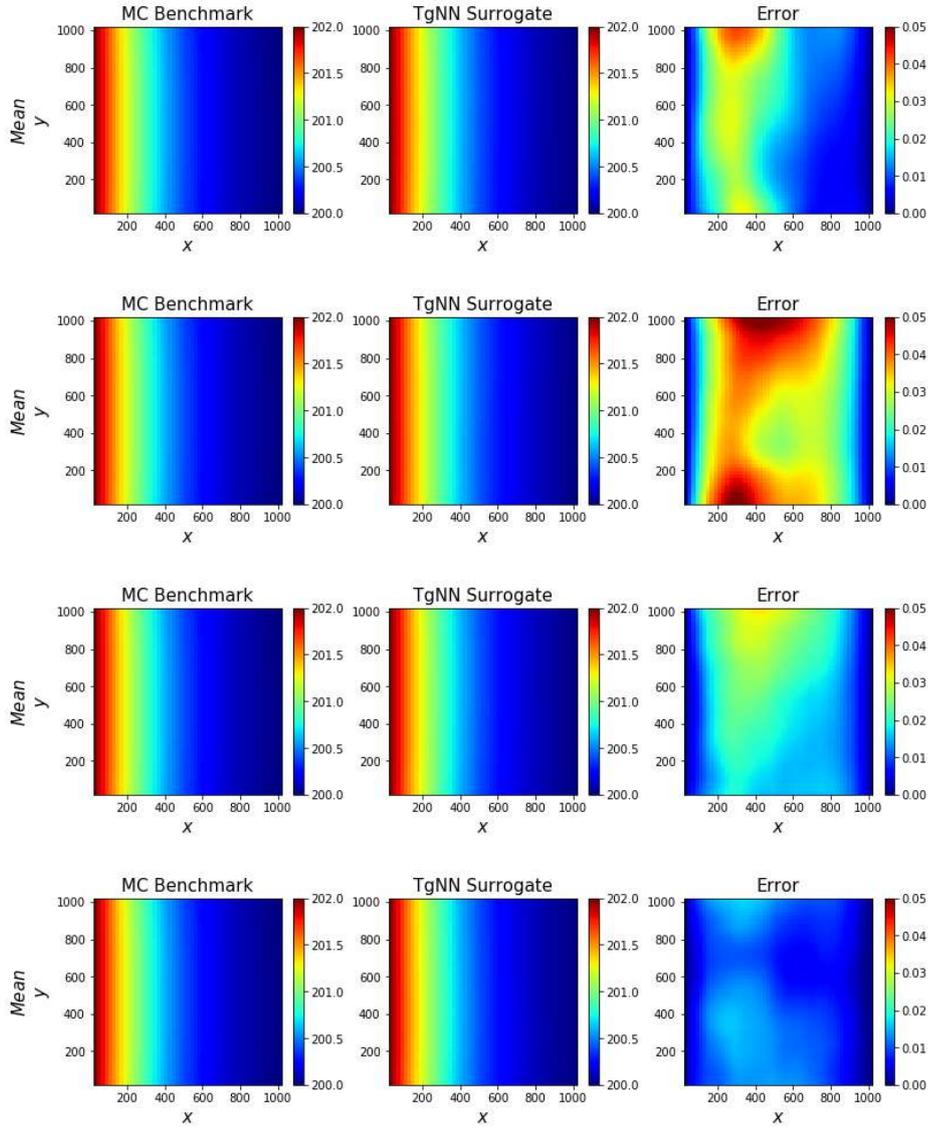

**Fig. A.1.** Estimated mean at time step 30 from different TgNN surrogates compared with the MC benchmark. The TgNN surrogates are trained with $1 \times 10^4$ (first row), $5 \times 10^5$ (second row), $1 \times 10^6$ (third row), and $1.5 \times 10^6$ (fourth row) collocation points, respectively.

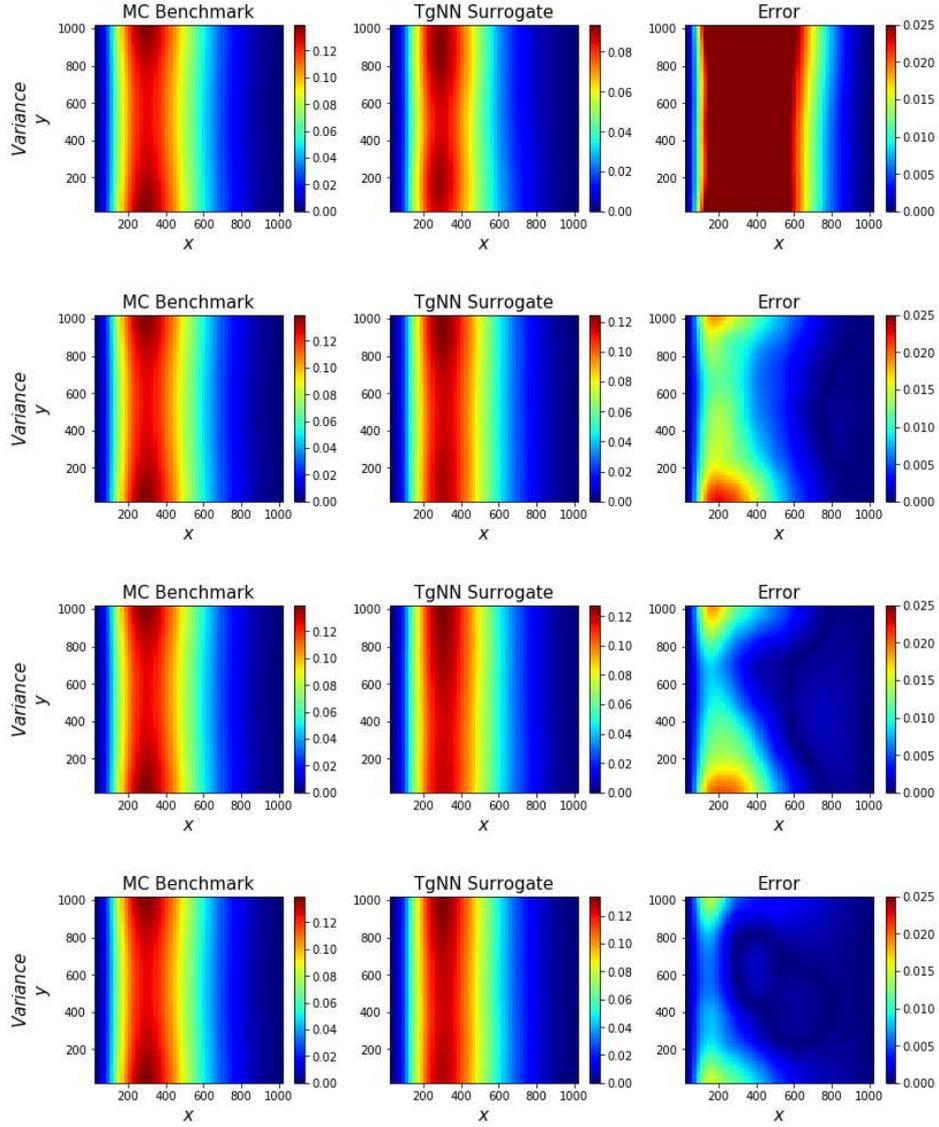

**Fig. A.2.** Estimated variance at time step 30 from different TgNN surrogates compared with the MC benchmark. The TgNN surrogates are trained with $1 \times 10^4$ (first row), $5 \times 10^5$ (second row), $1 \times 10^6$ (third row), and $1.5 \times 10^6$ (fourth row) collocation points, respectively.

**Appendix B**

In this appendix, the relative $L_2$ error and $R^2$ score of estimated statistical moments at different time steps for the base case from the TgNN surrogate are provided. The TgNN surrogate is trained with different numbers of collocation points ($N_c = 1 \times 10^5$, $N_c = 5 \times 10^5$, and $N_c = 1 \times 10^6$) and different numbers of labeled data ($R = 0$, $R = 1, R = 10$, and $R = 30$). The relative $L_2$ error and $R^2$ score of estimated mean and variance at different time steps from 10,000 realizations are shown in **Fig. B.1** and **Fig. B.2**, respectively. It can be seen that the TgNN surrogate can provide accurate UQ results at different times for dynamic subsurface flow problems when adequate labeled data and collocation points are employed.

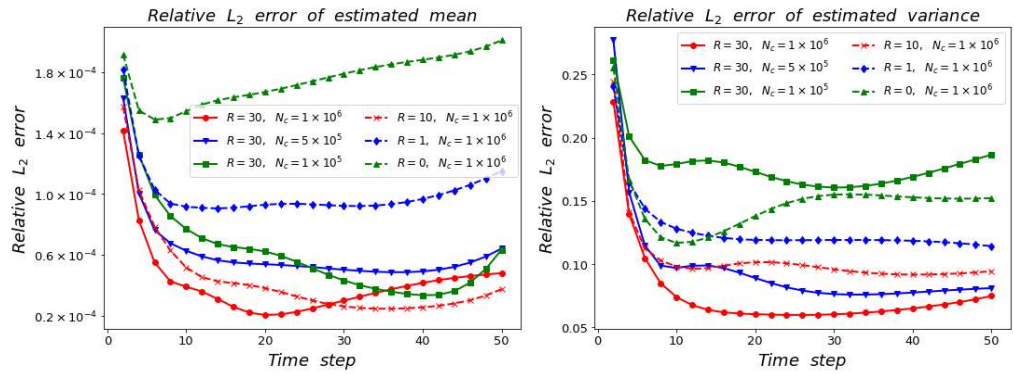

**Fig. B.1.** Relative $L_2$ error of estimated mean and variance at different time steps from different TgNN surrogates.

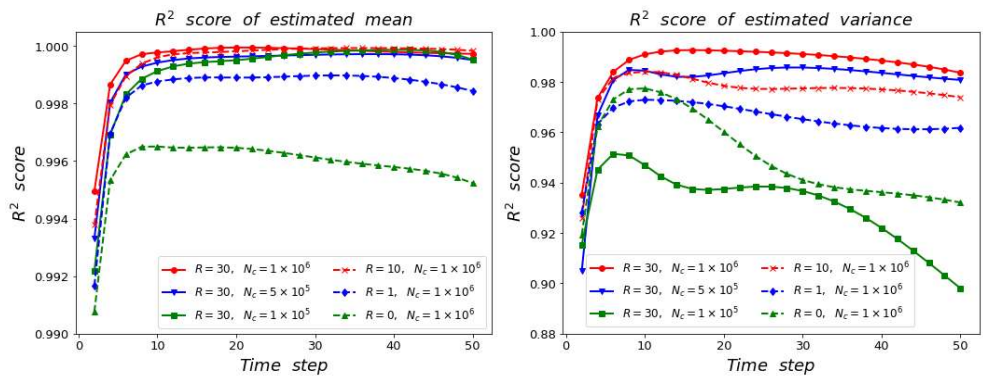

**Fig. B.2.** $R^2$ score of estimated mean and variance at different time steps from different TgNN surrogates.